\documentclass[11pt]{article}

\usepackage{arxiv}
\usepackage{amsmath}
\usepackage{subfigure}
\usepackage[utf8]{inputenc} 
\usepackage[T1]{fontenc}    
\usepackage{hyperref}       
\usepackage{url}            
\usepackage{booktabs}       
\usepackage{amsfonts,amssymb,amsbsy}       
\usepackage{mathtools}
\usepackage{lipsum}
\usepackage{enumitem}
\usepackage[linesnumbered]{algorithm2e}

\RestyleAlgo{ruled}

\usepackage{bm}

\title{A transport approach to sequential simulation-based inference}

\author{
  Paul-Baptiste Rubio \\
  Massachusetts Institute of Technology\\
  Cambridge, MA 02139 \\
  \texttt{rubiop@mit.edu} \\
   \And
 Youssef Marzouk\\
  Massachusetts Institute of Technology\\
  Cambridge, MA 02139 \\
  \texttt{ymarz@mit.edu} \\
     \And
 Matthew Parno\\
  Dartmouth College\\
  Hanover, NH 03755 \\
  \texttt{matthew.d.parno@dartmouth.edu} \\
  \texttt{} \\
}

\begin{document}
\setlength{\algomargin}{1.5em}

\maketitle

\begin{abstract}
We present a new transport-based approach to efficiently perform sequential Bayesian inference of static model parameters. The strategy is based on the extraction of conditional distribution from the joint distribution of parameters and data, via the estimation of structured (e.g., block triangular) transport maps. This gives explicit surrogate models for the likelihood functions and their gradients. This allow gradient-based characterizations of posterior density via transport maps in a model-free, online phase. This framework is well suited for parameter estimation in case of complex noise models including nuisance parameters and when the forward model is only known as a black box. The numerical application of this method is performed in the context of characterization of ice thickness with conductivity measurements.
\end{abstract}
\newpage

\section{Introduction}\label{sec:intro}

Bayesian inference is a popular way to perform robust parameter estimation in presence of uncertainties. To solve a Bayesian inference problem, one want to define a prior and likelihood function suited for the problem at hands. However, in some problems, likelihood and/or prior are unavailable or computationally prohibitive to evaluate. One use case is when considering nuisance parameters. In that case, evaluation of likelihood function requires integration over the potentially high dimensional nuisance parameter space making posterior sampling very difficult. In this context it might be easier to obtain joint samples of parameters and data from the simulator.

To solve Bayesian inference problems where only joint samples of parameter and data are available (or computationally easy to obtain), Simulation-Based Inference (SBI) methods also called Likelihood Free Inference (LFI) methods has emerged as a powerful tool for statistical inference in a wide range of applications \cite{Cranmer2020}. Simulation-based inference involves generating artificial data from a probabilistic model and comparing it to observed data to infer the underlying model parameters. Approximate Bayesian Computation (ABC) \cite{Rubin1984,Beaumont2002,Marjoram2003,Fearnhead2012} is the first class of techniques that allow characterization of the posterior without explicit knowledge of the likelihood. This technique is based on a rejection algorithm that select posterior candidates according to the distance between the corresponding model outputs and the observed data. This technique requires appropriate tuning of the discrepancy measure and is quite inefficient when samples need to be drawn according to posterior densities with i.i.d observations. To overcome these limitations amortized methods have been developed where model of likelihood or posterior are directly built. With the development of deep learning techniques \cite{DeepLearning} and using the approximation power of neural networks, amortized methods include Neural Ratio Estimation (NRE) \cite{Cranmer2015,Hermans2020,Thomas2022}, Neural Posterior Estimation (NPE) \cite{Papamakarios2016,Lueckmann2017,Zhegal2022} and Neural Likelihood Esitmation (NLE) \cite{Price2018,Frazier2022}. These methods have proven to be more effective than ABC for simulation-based inference, especially when dealing with high dimensional parameters and data.

These techniques allow to have a surrogate model of either: the likelihood ratio (NRE), the likelihood function (NLE) or directly the posterior (NLE). With those approximations, posterior samples can be obtained from the surrogate for different realization of the observations without re-training the models. While sampling from the posterior is the most direct approach to solve SBI problems, learning the likelihood is the most appropriate choice when the prior is not known or one wants to try different priors to solve a given set of inference problems. 

In the context of this paper, the choice of learning the likelihood function is motivated by solving sequential inference problems where the prior changes at each assimilation step. Solving sequential inference problems is required when observations are only available sequentially and that prediction need to be in real time (e.g., risk assessment) and actions on the system need to be performed (e.g., control) \cite{RubioControl}. In that case, full online characterization of posterior densities is required. 

In the previous work \cite{Rubio2019}, an approach has been proposed to solve online sequential inference problems. This approach requires the knowledge of a reduced model which allows pseudo-analytical posterior densities that can be characterized by the computations of transport maps \cite{Moselhy2011,Spantini2017}. In this paper it has been shown than efficient characterizations of sequential posterior densities is only made possible due to knowledge of gradient of the forward model with respect to the parameters. This method is quite intrusive as direct model evaluations are not sufficient to build the surrogate model. Recent works have been in the direction of building non-intrusive methods for uncertainty quantification and reduced order models in order to have a greater impact in the industry and deal with models only known as a "black-box" \cite{Herzog2008,Giraldi2014,Scanff2022}. These methods need to be adapted for each model/application and can't handle Bayesian inference problem including nuisance parameters or implicit noise. 

The approach proposed in this paper is fully non-intrusive, applicable for all types of black-box models and only requires joint samples of parameters and data. It consists in one measurement-free phase and one model evaluation-free phase. For the measurement-free phase, a stochastic surrogate model for the likelihood functions is computed. This is performed via the computation of triangular transport maps that exhibit the conditional density characterizing the likelihood function. This surrogate model can be seen as a function of the parameters of interest parameterized by  the observations variables. This allow to compute stochastic surrogate without the knowledge of the observations. This approach is then well suited for expensive models where the surrogate likelihood functions can be estimated offline and in parallel for each assimilation step. Then, for the model evaluation-free phase, posterior densities are characterized via the computation of a sequence of transport maps that allow easy sampling and computation of quantity of interest for the posterior densities. This lead to an online algorithm where the computation time per assimilation step is constant. 

The closest related work \cite{Papamakarios2018} uses normalizing flows to learn the likelihood function with samples drawn according to a sequence of intermediate posterior densities. In our work, for a given assimilation step, parameter samples are drawn according to the same distribution since the observation is considered to be unknown at the time of the training. Also, in this work we provide a specific posterior sampling method suited for sequential Bayesian inference instead of MCMC. Other work using transport maps for LFI include \cite{Spantini2019CouplingFiltering,Baptista2022LFI} where joint samples are used to transport standard normal samples to posterior samples.
The setting of this work differs from the setting of joint state-parameter estimation in dynamical systems treated by methods like particle MCMC \cite{AndrieuPMCMC} or SMC2 \cite{ChopinSMC2} which are computational heavy. Here state is automatically marginalized out and considered problems do not include stochastic dynamics. The presented approach  can however tackle parameter inference problems with deterministic dynamics plus implicit noise.

The present paper is organized as follow. Section \ref{sec:bayesian} presents setting of sequential Bayesian inference problem treated in this work. Details of how transport maps are built and computed are shown Section \ref{sec:tm_details} while the full algorithm for sequential simulation-based inference is presented in Section \ref{sec:ssbi_algorithm}. In Section \ref{sec:results}, in the context of sea ice thickness measurement, algorithm performances are studied on numerical examples of increasing complexities.

\section{Sequential Bayesian inference of static parameters}\label{sec:bayesian}
\subsection{Bayes formulation}\label{sec:static_bayes}

Bayesian inference is an extensively used approach to deal with parameters estimation when in presence of uncertainties. For a given set of parameters $\bm{\theta}\in\mathbb{R}^{n_\theta}$ and a given set of observations $\mathbf{y} \in \mathbb{R}^{n_y}$, the objective is to characterize the posterior density $\pi(\bm{\theta}|\mathbf{y})$. This conditional density can be expressed thanks to Bayes' formula as:
\begin{equation*}
    \pi(\bm{\theta}|\mathbf{y})\propto \pi(\mathbf{y}|\bm{\theta}).\pi(\bm{\theta})
\end{equation*}
The density $\pi(\bm{\theta})$ is the prior distribution that takes into account all the prior information we have about the distribution of parameters $\bm{\theta}$ without considering the information given by the observation $\mathbf{y}$. The other term in Bayes' formula is the likelihood function $\pi(\mathbf{y}|\bm{\theta})$. This term represents the conditional probability that the considered system outputs a quantity similar to the observation $\mathbf{y}$ for a fixed value $\bm{\theta}$. Hence, this function is seen as a function of $\bm{\theta}$ knowing the value of $\mathbf{y}$. The form of the likelihood function highly depends on the considered problem and the associated modeling choices. In this context, this work aims to consider the general case where the random variable $\bm{\epsilon}$, which represents noise in the system, can conditionally depends on parameter $\bm{\theta}$. This random variable follows the density $\pi_\epsilon(\bm{\epsilon}|\bm{\theta})$ and is related to the observable $\mathbf{y}$ as:
\begin{equation*}
    \bm{y}=H(\bm{\theta},\bm{\epsilon})
\end{equation*}
where, $H$ is the observation model.

Assuming there is no other source of uncertainty in the considered model, the probability density function of the observable given parameters and noise can be expressed with the Dirac delta function $\delta$:
\begin{equation} \label{eq:dirac}
    \pi(\mathbf{y}|\bm{\theta},\bm{\epsilon}) = \delta(\mathbf{y}-H(\bm{\theta},\bm{\epsilon}))
\end{equation}

From \eqref{eq:dirac}, noise can be marginalized out, and the likelihood function can be calculated as:
\begin{align}\label{eq:lkl_gen}
    \pi(\mathbf{y}|\bm{\theta}) &=\int \pi(\mathbf{y}|\bm{\theta},\bm{\epsilon}).\pi_\epsilon(\bm{\epsilon}|\bm{\theta}) \text{d} \bm{\epsilon} \\
    &= \int \delta(\mathbf{y}-H(\bm{\theta},\bm{\epsilon})).\pi_\epsilon(\bm{\epsilon}|\bm{\theta}) \text{d} \bm{\epsilon}
\end{align}

When facing an additive noise model, i.e.:
\begin{equation}\label{eq:add_noise}
    \mathbf{y} = H(\bm{\theta}) + \bm{\epsilon},
\end{equation}
the likelihood in \eqref{eq:lkl_gen} can be expressed as:
\begin{equation}\label{eq:lkl_add}
    \pi(\mathbf{y}|\bm{\theta}) = \pi_\epsilon(\mathbf{y}-H(\bm{\theta})),
\end{equation}Similarly, an explicit form can be found when dealing with the following multiplicative noise:
\begin{equation*}
\mathbf{y} = H(\bm{\theta}).\bm{\epsilon}, \,\,\, H(\bm{\theta}) \neq 0
\end{equation*}In that case, likelihood function reads:
\begin{equation*}
\pi(\mathbf{y}|\bm{\theta})= \frac{1}{H(\bm{\theta})}.\pi_\epsilon \left(\frac{\bm{y}}{H(\bm{\theta})}\right)
\end{equation*}

With other more complex cases, likelihood functions have no explicit closed forms. One objective of this work is to consider this setting of complex noise model. This context includes situations where only parameters and corresponding noisy system outputs samples are available. An other situation of interest is when the system is surrounded by aleatory variables/nuisance parameters $\bm{\eta}$ (opposed to epistemic variables $\bm{\theta}$). In that case noise can be decomposed as:
\begin{equation*}
    \bm{\epsilon} = (\bm{\xi},\bm{\eta}) \,\, \text{with } \,\, \bm{\xi} \perp \!\!\! \perp \bm{\eta}
\end{equation*}
Hence, the likelihood function reads:
\begin{equation}\label{eq:general_lkl}
    \pi(\mathbf{y}|\bm{\theta}) =\int \int \delta(\mathbf{y}-H(\bm{\theta},\bm{\xi},\bm{\eta}).\pi_\xi(\bm{\xi}|\bm{\theta}).\pi_\eta(\bm{\eta}|\bm{\theta}) \text{d} \bm{\xi}\text{d} \bm{\eta}
\end{equation}
Then, even if the noise model is explicit (e.g, additive or multiplicative) and that the likelihood function $\pi(\mathbf{y}|\mathbf{\theta},\bm{\eta})$
is then known in closed form, a marginalization with respect to $\bm{\eta}$ is required, i.e.,
\begin{equation} \label{eq:lkl_nuis}
        \pi(\mathbf{y}|\bm{\theta}) =\int \pi(\mathbf{y}|\bm{\theta},\bm{\eta}).\pi_\eta(\bm{\eta}|\bm{\theta}) \text{d}\bm{\eta}
\end{equation}
The difference between $\bm{\theta}$ and $\bm{\eta}$ is based on modeling choice and/or prior assumptions. According to \cite{basu}:
\textit{If the inference problem at hand relates only to $\bm{\theta}$ and if information gained on $\bm{\eta}$ is of no direct relevance to the problem, then we classify $\bm{\eta}$ as the nuisance parameter}. In Bayesian inference it assumed that gathering data doesn't help reducing the uncertainty on the parameters. 

It is worth noting that when dealing with nuisance parameters, likelihood functions and therefore Bayesian formulation can be difficult to formulate/exploit due to the required integration in \eqref{eq:lkl_nuis}. One objective of this paper is to efficiently characterize posterior densities define with these general type of likelihood functions.

\subsection{Sequential setting}\label{sec:sequential_bayes}

In the sequential setting with assume that a new observation $\mathbf{y}_t$ is available at each time step $t$. The purpose is then to characterize the posterior distribution $\pi(\bm{\theta}|\mathbf{y}_{1:t})$ at each time step $t$. The most general formulation of the Bayesian posterior for this problem reads:
\begin{equation*}
    \pi(\bm{\theta}|\mathbf{y}_{1:t-1},\mathbf{y}_{t}) \propto \pi(\mathbf{y}_{t}|\bm{\theta},\mathbf{y}_{1:t-1}).\pi(\bm{\theta}|\mathbf{y}_{1:t-1})
\end{equation*}
This formulation can be interpreted as the posterior at time step $t$ being proportional to the product of the previous posterior at time step $t-1$ and a new likelihood which, in addition to parameters, is also conditioned on the previous $t-1$ observations. 

With some assumptions, the likelihood function for sequential parameter estimation can be simplified in product of likelihood functions defined as in section \ref{sec:static_bayes} for each time step and the first prior.
In the particular case where the nuisance parameters $\bm{\eta}$ can be decomposed as independent nuisance parameters for each time step and that each of nuisance parameters are statistically independent of previous observations, the likelihood function does not depend on previous observations, i.e.:
\begin{equation}\label{hyp:nuisance}
    \pi(\mathbf{y}_{t}|\bm{\theta},\mathbf{y}_{1:t-1})=\pi(\mathbf{y}_{t}|\bm{\theta})\text{, for } \pi(\bm{\eta})=\pi(\bm{\eta}_1) \cdots \pi(\bm{\eta}_t)\text{ and } \mathbf{y}_k|\boldsymbol{\theta}  \perp \!\!\! \perp \{ \boldsymbol{\eta}_i \}_{i \neq k}  
\end{equation}

In that case, likelihood functions at each time step $t$ read:
\begin{equation}
        \pi(\mathbf{y}_t|\bm{\theta}) =\int \pi(\mathbf{y}_t|\bm{\theta},\bm{\eta}).\pi_{\eta_t}(\bm{\eta}_t|\bm{\theta}) \text{d}\bm{\eta}_t
\end{equation}

Hence the Bayesian formulation reads:
\begin{equation}\label{eq:seq_bayes}
    \pi(\bm{\theta}|\mathbf{y}_{1:t}) \propto \pi(\mathbf{y}_{t}|\bm{\theta}).\pi(\bm{\theta}|\mathbf{y}_{1:t-1})
\end{equation}

The following work will be based on the hypothesis \eqref{hyp:nuisance}. For other forms of nuisance parameters, the formulation \eqref{eq:seq_bayes} will only consists in an approximation. Example with this hypothesis will be treated in Section \ref{sec:results}. 

In this context, deriving tractable forms of likelihood functions and tractable sampling methods for the posterior density are very challenging. This becomes even more challenging when trying to design online algorithms (with constant cost per time step) to perform sequential Bayesian inference. To tackle those challenges an approach based on the computation of triangular transport maps is detailed in the following.

\section{Triangular transport maps for conditional density estimation and sampling} \label{sec:tm_details}

\subsection{Triangular transport maps}\label{sec:triangular_maps}
Considering two probability density functions $\rho$ and $\pi$, the objective is to construct an invertible application $S:\mathbb{R}^d\rightarrow\mathbb{R}^d$ such that knowing one density and this map will lead to knowing the other density via the change of variable formula:
\begin{equation}\label{eq:pullback}
   \pi(\mathbf{x}) = S^\sharp \rho\left(\mathbf{x}\right) = \rho(S\left(\mathbf{x}\right))|\det \nabla S\left(\mathbf{x}\right) |
\end{equation}
$S^\sharp \rho(\mathbf{x})$ and $._\sharp$ are respectively called pullback density and pullback operator. We can equivalently define the pushforward density via the transport map $T$ as:
\begin{equation}\label{eq:pushforward}
\pi(\mathbf{x}) = S_\sharp \rho(\mathbf{x}) = \rho\left(T^{-1}\left(\mathbf{x}\right)\right)|\det \nabla T^{-1}\left(\mathbf{x}\right) |
\end{equation}

Practically, the objective is to choose a reference density $\rho$ (e.g., standard normal) and compute a map that pushes forward or pulls back $\rho$ to $\pi$ knowing some information about $\pi$ (samples or unnormalized pdf). Once this map is found, by applying \eqref{eq:pullback} or \eqref{eq:pushforward} an approximation of $\pi$ can be made (density approximation). An other application of the transport map is for sampling. Indeed, knowing samples $\mathbf{x}^i$ from $\rho$, samples from $\pi$ are $S^{-1}(\mathbf{x}^i)$ in the case of \eqref{eq:pullback} or $T(\mathbf{x}^i)$ in the case of \eqref{eq:pushforward}.

In this work, transport maps are restricted to be lower triangular. This means that each component $S_k:\mathbb{R}^k\rightarrow\mathbb{R}$ only depends on the first $k$-th variables $\mathbf{x}_{1:k}=(x_1,...,x_k)\in\mathbb{R}^k$, i.e.,:
\begin{equation*}
    {S}(\mathbf{x})=
    \begin{bmatrix*}[l]
        {S}_1(x_1)\\
        {S}_2(x_1,x_2)\\
        \,\,\,\,\,\,\vdots\\
        {S}_d(x_1,x_2,...,x_d)
      \end{bmatrix*}
\end{equation*}

The choice of this particular structure known as the \textit{Knothe-Rosenblatt} (KR) rearrangements is motivated by several properties:
\begin{itemize}
    \item The transport map exits and is unique under mild assumptions on the target $\pi$ and reference $\rho$
    \item Invertibility of the map is guaranteed by one-dimensional monotonicity $\partial_k S_k>0$
    \item Computation of the inverse $S^{-1}$ consists in solving $d$ one-dimensional root finding problems.
    \item The term $\det | \nabla S (\mathbf{x}) |$ is simple to evaluate
    \item A specific ordering of the variables $(x_1,x_2,...,x_d)$ allow to decompose the joint density as:
    \begin{equation*}
        \pi(\mathbf{x})=\pi(x_1)\pi(x_2|x_1)...\pi(x_d|x_{1:d-1})
    \end{equation*}
where each conditional density $\pi(x_k|x_{1:k})$ is characterized by map component $S_k$.
\end{itemize}

\subsection{Monotone parametrization}
In order to make the problem of finding triangular maps computationally tractable the problem is discretized by using a parametrization of map components and solving a variational problem. Here, map component are parametrized with multivariate polynomials. To enforce the monotonicity constraint $\partial_k S_k>0$ , following the approach detailed in \cite{BaptistaATM}, the operator $\mathcal{R}_k$ that transforms any sufficiently smooth basis function $f$ into a function increasing in the $k$-th variable is used. This operator, called rectifier is defined as:
\begin{equation}\label{eq:rectifier}
\mathcal{R}_k(f) (\mathbf{x}_{1:k}) = f(\mathbf{x}_{1:k-1},0)+\int_{0}^{x_k} g(\partial_k f(\mathbf{x}_{1:k-1},t)) \text{d}t,
\end{equation}
where $g:\mathbb{R}\rightarrow\mathbb{R}^*_+$ is a strictly positive function. One common choice is to choose $g$ as the softplus function i.e., $g(x)=\log(1+\exp(x))$.

\subsection{Objective functions}

With the monotone paramterization defined in \eqref{eq:rectifier}, different ways are possible in order to find the optimal base function $f$ and hence map $S$ that pushes reference density $\rho$ to the target $\pi$. The different ways of finding $S$ depend on the information available.
\subsubsection{Map from density}
When the information available about the target $\pi$ is the expression of the unnormalized probability density function $\bar{\pi}$, the objective function is based on the minimization of the Kullback-Liebler (KL) divergence between $S_\sharp \rho$ and $\pi$ with respect to the map $S$. 
\begin{align}
    D_{KL}(S_\sharp \rho || \pi)&=D_{KL}(\rho ||S^\sharp \pi) \\
    &=\mathbb{E}_\rho \left[ \log(\rho) - \log(S^\sharp \pi) \right] \label{eq:KL_density}
\end{align}
Since the goal is to find minimize the KL divergence with respect to $S$, term $\log(\rho)$ can be removed from the objective function based on \eqref{eq:KL_density}. For the same reason, to build the objective function density $\pi$ only need to be known up to a constant meaning we can replace $\pi$ by its unnormalized form $\bar{\pi}$ for the objective function. With those argument the continuous objective function for the map from density case reads:
\begin{equation} \label{eq:cont_kl_dens}
    J(S) = - \mathbb{E}_\rho\left[ \log(S^\sharp \bar{\pi} ) \right]
\end{equation}

In general, expectation in \eqref{eq:cont_kl_dens} can't calculated in closed form. In practice the discrete form of this objective is used when knowing samples $\mathbf{x}^i\sim \rho$, $i\in\{1,...,n\}$:
\begin{equation}\label{eq:discr_kl_dens}
    \hat{J}(S) = - \frac{1}{N} \sum_{i=1}^n \log(S^\sharp \bar{\pi}(\mathbf{x}^i )), \,\, \mathbf{x}^i \sim \rho, \,\, i \in \{1,...,n\}
\end{equation}

\subsubsection{Map from samples}\label{sec:map_from_samples}
When samples $\mathbf{x}^i$, $i \in \{1,...,n\}$ from $\pi$ is the information at hand, the KL divergence from where the objective function is derived reads:
\begin{equation}
    D_{KL}(\pi || S^\sharp \rho)=\mathbb{E}_\pi \left[ \log(\pi) - \log(S^\sharp \rho) \right] \label{eq:KL_samples}
\end{equation}
Similarly to the map from density case, term $\log(\pi)$ in \eqref{eq:KL_samples} can be removed to build the objective function and discrete form of the expectation can be used since we have samples from $\pi$:
\begin{equation}\label{eq:discr_kl_samp}
    \hat{J}(S) = - \frac{1}{n} \sum_{i=1}^N \log(S^\sharp {\rho}(\mathbf{x}^i )), \,\, \mathbf{x}^i \sim \pi, \,\, i \in \{1,...,n\}
\end{equation}
One additional property of the objective function in \eqref{eq:discr_kl_samp} comes when $\rho$ is chosen as a product density i.e., $\rho(\mathbf{x}_{1:d})=\rho_1(x_1)\times...\times\rho_d(x_d)$ (e.g. when using the standard normal Gaussian). In that case, the objective function can be decomposed in $d$ independent objective functions on the $d$ map components $S_k$, $k \in \{1,...,d\}$. Those objective functions read:
\begin{equation}\label{eq:discr_kl_samp_k}
    \hat{J}_{k}(S_k) = - \frac{1}{n} \sum_{i=1}^n \log(S_k^\sharp {\rho_k}(\mathbf{x}_{1:k}^i )), \,\, \mathbf{x}^i \sim \pi, \,\, i \in \{1,...,n\},\,\, \forall k \in \{1,...,d\}.
\end{equation}
This decomposition is highly beneficial because it leads to smaller problems to solve that can be solved independently in parallel.

\subsubsection{Map from regression}
The last configuration addressed in this work is when map evaluations $\mathbf{z}^i$, at points $\mathbf{x}^i$ are directly available. In that case, to fit a map $S$ to those points a simple least-square objective can be used. As in the map from samples case, $d$ independent objective functions can be derived and read:
\begin{equation}\label{eq:regression_obj}
    \hat{J}_k(S_k) = \frac{1}{2} \sum_{i=1}^N ( S_k(\mathbf{x}_{1:k}^i) - y_k^i)^2, \,\, \forall k \in \{1,...,d\}.
\end{equation}

\subsection{Adaptive procedure} \label{sec:atm}
\subsubsection{Multivariate polynomial expansion}
With the definition of the rectifier in \eqref{eq:rectifier}, finding a suitable transport maps ends up finding optimal function $f$. In this work base functions $f$ are parameterized with multivariate polynomial expansions i.e,:
\begin{equation*}
    f(\mathbf{x}_{1:d})=\sum_{\alpha \in \mathcal{A}}\bm{w}_\alpha \Phi(\mathbf{x}_{1:d})
\end{equation*}
where $\alpha$ are multi-indices in the set $\mathcal{A}$ that define the terms in the expansion (e.g., the order of each polynomial term), $\bm{w}_\alpha$ are coefficients and $\Phi(\mathbf{x}_{1:d})$ are multivariate basis functions defined through the tensor product of one-dimensional polynomials:
\begin{equation*}
\Phi_i(\mathbf{x}_{1:d})=\prod_{i=1}^d \phi_{\alpha_i}(x_i)
\end{equation*}
For example a common choice is to choose $\phi_{\alpha_i}$ to be Hermite polynomials of order $\alpha_i$.

With this parameterization objective functions \eqref{eq:discr_kl_dens}, \eqref{eq:discr_kl_samp_k} and \eqref{eq:regression_obj} can be transposed to objective function over the vector of polynomial coefficients $\bm{w}_k$ that define base functions $f$ in the rectifier definition in \eqref{eq:rectifier}.

\subsubsection{Adaptive selection of multi-indices}
Using the rectifier operator with polynomial approximation of base functions lead to cheap and easy way to parameterize transport maps. However, increasing the dimension $d$ of the problem lead to an exponential growth of number of polynomial basis and hence number of parameters (coefficients) to optimize. While increasing the order of polynomials allows to improve map expressiveness, this could lead to intractable computations when considering increasing dimensional problems. This problem also called "curse of dimensionality" has recently been addressed in \cite{BaptistaATM} for the computation of map from samples. The benefits of the Adaptive Transport Maps (ATM) approach presented in this paper are many. First, it allows to precisely increase the complexity of the maps allowing small adaptation steps in order to avoid the curse of dimensionality. Second, this approach allows to adapt and discover sparse structure in considered problems. Finally the ATM approach gives stopping criteria for the adaptation procedure which leads to finding the transport maps with the best bias-variance trade-off using a fixed number of training samples.

The adaption procedure starts with a downward closed multi-index set $\mathcal{A}$ and look at candidate multi-indices $\alpha \in \mathcal{A}^\text{RM}$, where $\mathcal{A}^\text{RM}$ is the reduced margin. Then, the best candidate $\alpha^*$ is added to $\mathcal{A}$ according to the criterion of best objective function improvement:
\begin{equation}\label{eq:argmax_alpha}
    \alpha^* = \underset{\alpha}{\text{arg max}} \left| \nabla_\alpha \hat{J}_k(S_k) \right|
\end{equation}

In practice this is done by computing the gradient of the objective function with respect to coefficients $\bm{w}_\alpha$ and $\alpha^*$ is 
 the multi-index corresponding to coefficient $\bm{w}_{\alpha^*}$ which maximizes the objective function.

While this approach has been designed for the objective function \eqref{eq:discr_kl_samp_k}. In this work this adaptive procedure has been extended to objective functions \eqref{eq:discr_kl_dens} and \eqref{eq:regression_obj} to solve corresponding map estimation problems.

\subsubsection{Stopping criteria}
One last import key of the adaptation algorithm is the stopping criterion. Having an appropriate stopping criteria is of a primary importance when the maps are used in a context of real-time simulation. In \cite{BaptistaATM}, a cross-validation criteria is proposed based on a held out validation sample set. Such a validation set based criterion is necessary for objective function \eqref{eq:discr_kl_samp_k} since its exact minimum is unknown. 

For the map from density problem, other stopping criteria can be derived. The first one is the so-called variance diagnostic $\epsilon_\sigma$ defined in \cite{Moselhy2011} as:
\begin{equation*}
    \epsilon_\sigma(S_\sharp \rho,\pi) = \frac{1}{2} \mathbb{V}\text{ar} \left[\log \frac{\rho}{S^\sharp\bar{\pi}} \right] \approx D_{KL}(S_\sharp \rho || \pi)
\end{equation*}

This diagnostic computed thanks to testing samples (independent of training samples) from the reference and the knowledge of the unnormalized density $\bar{\pi}$ is a direct approximation of the KL divergence involved in the objective \eqref{eq:cont_kl_dens}. Its convergence to zero can be monitored to define a stopping criteria. In is discrete from the variance diagnostic used in practice reads:
\begin{equation}\label{eq:var_diag_discr}
    \hat{\epsilon}_\sigma(S_\sharp \rho,\pi) = \frac{1}{2 n_\text{test}} \sum_{i=1}^{n_\text{test}} \left[\left(\log\bar{\pi}(\textbf{x}^i) - \log S_\sharp\rho(\textbf{x}^i) \right) - \frac{1}{n_\text{test}} \sum_{i=1}^{n_\text{test}} \left(\log\bar{\pi}(\textbf{x}^i) - \log S_\sharp\rho(\textbf{x}^i) \right) \right]^2, \,\, \textbf{x}^i \sim \rho
\end{equation}

The second stopping criteria for the map from density problem is a criteria derived from an approximation of an other KL divergence. Indeed, as it has been proven in \cite{Brennan2019}, the following holds:
\begin{equation}\label{eq:trace_diag_cont}
    D_{KL}(\pi || S_\sharp \rho) \leq \frac{1}{2} \text{trace}\left[ \int \left(\nabla \log \frac{S^\sharp \bar{\pi}}{\rho} \right)\left(\nabla \log \frac{S^\sharp \bar{\pi}}{\rho} \right)^T \text{d}S^\sharp\pi \right]
\end{equation}

Since no samples from $\pi$ are known, integral in \eqref{eq:trace_diag_cont} can be approximated with samples from the reference $\rho$. Hence, the discrete approximated form of the trace diagnostic that will be use in this work reads:
\begin{equation} \label{eq:trace_diag_discr}
    \hat{\epsilon}_\text{trace}(\pi,S_\sharp \rho) = \frac{1}{2 n_\text{test}} \text{trace}\left[\sum_{i=1}^{n_\text{test}} \left(\nabla \log \frac{S^\sharp \bar{\pi}(\bm{x}^i)}{\rho(\bm{x}^i)} \right)\left(\nabla \log \frac{S^\sharp \bar{\pi}(\bm{x}^i)}{ \rho(\bm{x}^i)} \right)^T \right], \,\, \textbf{x}^i \sim \rho
\end{equation}

This diagnostic is also computed with held-out testing samples from $\rho$ and only requires the evaluation of the unnormalized density $\bar{\pi}$. This diagnostic also converges to zero as density $S_\sharp \rho$ is closer to $\pi$.  

One final criterion used in this work is particular to the estimation of maps with regression. Since, the objective function in \eqref{eq:regression_obj} is known to be zero at convergence, one easy stopping criterion that can be defined is simply a threshold value on the objective function evaluated at held-out testing samples.

While for the computation of maps from samples the procedure used in this work is exactly the same than described in \cite{BaptistaATM}, ATM has been adapted for all other objective functions and coupled with corresponding specific stopping criteria. The use of stopping criteria based on threshold values is of great benefits compared to the use of the stopping criterion based on the stagnation of the objective function. The first benefit is that it will in general provide earlier stopping. The second benefit is that since the stopping criteria are based on quantities converging to zero, at the end of each adaptation procedure these will give an information of how good the map approximation is. All of this will be even more beneficial when the map computations have to be performed with limited time resources which is one objective of this work.

\section{Sequential Bayesian inference algorithm}\label{sec:ssbi_algorithm}
The main objective of the algorithm is to provide characterizations of the full posterior densities in the framework of sequential Bayesian inference described in \eqref{eq:seq_bayes}. This framework includes cases where the model is known as a black box, is expensive, includes nuisance parameters and/or when online time resources are limited. To solve problems formulated within this framework a transport-based algorithm in two phases is proposed in the following.

\subsection{Phase I: building surrogate likelihood functions}
In the first phase, the objective is to approximate the likelihood functions $\pi(\mathbf{y}_t|\bm{\theta})$ involved in the posterior density written in \eqref{eq:seq_bayes}. In order to achieve that, samples $(\mathbf{y}_t^i,\bm{\theta}^i)$, $i \in \{1,...,n\}$ from the joint density $\pi(\bm{\theta},\mathbf{y}_t)$ can be generated as follow:
\begin{enumerate}
    \item Draw parameter samples from a candidate distribution $\pi_0$ (e.g., the prior):
    \begin{equation*}\label{eq:param_samps}
        \bm{\theta}^i \sim \pi_0(\bm{\theta})
    \end{equation*}
    \item Draw samples from measurement noise and nuisance parameters:
    \begin{equation*}
        \bm{\xi}_t^i \sim \pi(\bm{\xi}_t), \bm{\eta}_t^i \sim \pi(\bm{\eta}_t)
    \end{equation*}
    \item Evaluate the corresponding outputs using the black box model $H_t$:
    \begin{equation}\label{eq:out_samps}
        \bm{y}_t^i = H_t(\bm{\theta}^i,\bm{\xi}_t^i,\bm{\eta}_t^i)
    \end{equation}
    Model $H_t$ depends on some sequential parameter $t$ (e.g., time steps). The dependency of this model with respect to $t$ can be explicit or not and $\bm{y}_t^i$ are model output trajectories.
\end{enumerate}

Hence, quantities in \eqref{eq:param_samps} and \eqref{eq:out_samps} are samples from joint density $\pi(\bm{\theta},\mathbf{y}_t)$ while measurement noise and nuisance parameters samples are marginalized out without cost simply by being "thrown out".

The next step of this phase is to compute the triangular transport map $S$ that pulls back the reference density $\rho$ to the joint density $\pi(\bm{\theta},\mathbf{y}_t)$. This is done using the method described in \ref{sec:map_from_samples} and the ATM algorithm originally described in \cite{BaptistaATM}. Once the map $S$ is found the following approximation holds:
\begin{equation*}
    S^\sharp \rho (\bm{\theta},\mathbf{y}_t) \approx \pi(\bm{\theta},\mathbf{y}_t)
\end{equation*}
with:
\begin{equation*}
    {S}(\bm{\theta},\mathbf{y}_t)=
    \begin{bmatrix*}[l]
        {S}_{\bm{\theta}}(\bm{\theta})\\
        {S}_{\mathbf{y}_t}(\bm{\theta},\mathbf{y}_t)
      \end{bmatrix*}
\end{equation*}
With that particular ordering of variables and the properties of triangular map exposed in \ref{sec:triangular_maps}, the lower block ${S}_{\mathbf{y}_t}$ characterizes the likelihood function $\pi(\mathbf{y}_t|\bm{\theta})$ for all values of $\mathbf{y}_t$ and 
$\bm{\theta}$:
\begin{equation*}
    {S}_{\mathbf{y}_t}(\bm{\theta};.)^\sharp \rho(\mathbf{y}_t) \approx \pi(\mathbf{y}_t|\bm{\theta})
\end{equation*}

This method allows to compute surrogate likelihood functions in the most general setting \eqref{eq:general_lkl}. In this context, nuisance parameters can easily be marginalized out, even in large dimension, just by ignoring the samples when computing the map $S$.

It is worth noting that a similar method can be used to directly characterize the posterior $\pi(\mathbf{y}_t|\bm{\theta})$ just by inverting the ordering of the variables $\mathbf{y}_t$ and $\bm{\theta}$. This idea is used in the transport based data assimilation approaches \cite{Spantini2019CouplingFiltering,Baptista2022LFI} where lower block of joint triangular transport maps are computed to directly characterize posterior densities. However, learning the likelihood function instead of the posterior has the benefit of not involving the knowledge of the prior $\pi(\bm{\theta}|\mathbf{y}_{t-1})$. Indeed, in the process described above only samples from a proposal $\pi_0$, are required to generate samples used to build the transport. This proposal density can be any density close enough of the posterior density that shares the same support. In \cite{Papamakarios2018} this assumption on the proposal is even more relaxed as it is proven than any couple $(\bm{\theta}^i,\mathbf{y}^i)$ where $\bm{\theta}^i$ can come from different densities for each sample $i$ and where $\mathbf{y}^i$ is the corresponding output. 

In this phase, computation of map $S_{\mathbf{y}_t}$ is used to approximate the likelihood function for all possible values of observations making this phase "measurement-free". With that property, computation  be performed for assimilation step $t$ asynchronously from the real-time (when the observation is measured). This allows to have a functional solution of the sequential Bayesian problem with the dependency of the posterior density with respect to the observation which is known \textit{a priori}.

In addition to a closed form approximation of the likelihood function, knowing map $S_{\mathbf{y}_t}$ also gives access to a closed form representation of the gradient log-likelihood with respect to parameters $\bm{\theta}$:
\begin{equation*}
    \nabla_{\bm{\theta}} \log \pi(\mathbf{y}_t|\bm{\theta}) = \nabla_{\bm{\theta}}S_{\mathbf{y}_t}(\bm{\theta},\mathbf{y}_t).\nabla_{\mathbf{y}_t} \log \left[  S_{\mathbf{y}_t}(\bm{\theta},.)^\sharp \rho(\mathbf{y}_t)\right],
\end{equation*}
leading to also know a closed form approximation of the gradient log-posterior with respect to parameters.

Summary of computations required for building surrogate likelihood functions is described in Algorithm \ref{alg:offline_phase}.
\begin{algorithm}
\caption{Computation of surrogate likelihood functions from samples}\label{alg:offline_phase}

\For{$t \in \{1,...,n_t\}$}{
\For{$i \in \{1,...,n\}$}{
 Generate parameter samples: $\bm{\theta}^i \sim \pi_0(\bm{\theta})$
 
 Generate noise samples: $\bm{\xi}_t^i \sim \pi(\bm{\xi}_t)$ 

 Generate nuisance samples: $\bm{\eta}_t^i \sim \pi(\bm{\eta}_t)$ 

 Compute model outputs: $\mathbf{y}_t^i = H_t(\bm{\theta}^i,\bm{\xi}_t^i,\bm{\eta}_t^i)$
 }
Compute $S_{\mathbf{y}_t}$ such that: $S_{\mathbf{y}_t}(\bm{\theta};.)^\sharp \rho (\mathbf{y}_t)=\pi(\mathbf{y}_t|\bm{\theta})$ with samples $(\bm{\theta}^i,\mathbf{y}_t^i)$.
 }
\end{algorithm}

\subsection{Phase II: characterizing posterior densities}
From the previous phase, closed form of posterior density approximations can be known \textit{a priori}. In the online phase (when the measurements are actually known) only characterization of the posterior density (e.g., sampling) is needed to solve the Bayesian inference problem. Since the probability density function of the posterior is known explicitly at each time step, any sampling method like MCMC could be used at this stage. However, since the problem is sequential, a more adapted/faster method needs to be used in order to perform real-time online computations. For this purpose, in this work, posterior characterizations are made sequentially using composition of intermediate transport maps.

\subsubsection{Composition of intermediate transport maps}
The approach is similar to the one used in \cite{Rubio2019} and based on the initial work \cite{Spantini2017}. The idea is to find a transport map from the reference $\rho$ to each posterior $\pi(\bm{\theta}|\mathbf{y}_t)$, for each time step $t$. To do so, intermediate maps $T_t$ between consecutive posteriors are computed at each time step and the posterior map pushing forward $\rho$ to $\pi(\bm{\theta}|\mathbf{y}_t)$ is defined by composition as $\mathcal{T}_t=T_1 \circ ... \circ T_t$. Intermediate maps are defined as follow:
\begin{itemize}
    \item For the first time step:
\begin{equation}\label{eq:target_interm1}
    {T_1}_\sharp \rho = \pi(\bm{\theta}|\mathbf{y}_1),
\end{equation}
and $\mathcal{T}_1 = T_1$.
\item For time steps $t>1$:
\begin{align}
{T_t}_\sharp \rho &= \mathcal{T}_{t-1}^\sharp \pi(\bm{\theta}|\mathbf{y}_{1:t}) \\
\Rightarrow {T_t}_\sharp \rho &= \pi(\mathbf{y}_t|\mathcal{T}_{t-1} (\bm{\theta}))\rho(\bm{\theta}) 
\label{eq:target_interm2}
\end{align}
and $\mathcal{T}_t = \mathcal{T}_{t-1} \circ T_t$.
\end{itemize}
Computing intermediate maps instead of direct maps between reference and posteriors is advantageous for a couple of reasons. First, the target density $\pi(\mathbf{y}_t|\mathcal{T}_{t-1}(\bm{\theta}))\rho(\bm{\theta})$ is cheaper to evaluate than the product of the $t$ likelihood functions which composed $\pi(\bm{\theta}|\mathbf{y}_{1:t})$ since the information about all the previous observations is embedded in the map $\mathcal{T}_{t-1}$ that is not re-computed at time step $t$. Second, intermediate maps are in general easier to compute than direct maps. Indeed, the closer two densities are the simpler the map that transport one to another is. Then, depending on the sequential problem at hands, small variations between two consecutive posteriors can be expected. 

In order to approximate transport map from $\rho$ to target densities defined in \eqref{eq:target_interm1} and \eqref{eq:target_interm2}, map from density described in \ref{sec:triangular_maps}. This is has been made possible thanks to previous phase where closed form of the likelihood function and gradient of the log-likelihood functions have been derived. This allows the use of gradient based optimizer to find the maps and the gradient information also allows to the ATM procedure which used a criterion where gradient information of the log-posterior density is required (see \eqref{eq:argmax_alpha}).

\subsubsection{Map compression}
The use of composition of intermediate maps to characterize sequential posteriors can however have the inconvenient to be more and more expensive to evaluate as long as the number of assimilation time step growth. To circumvent this issue, when the composition of maps is considered to be too large, the whole composition of maps is replaced by one map obtained via regression. The idea is to compute the map ${\mathcal{T}}_t$ that approximate the map composition $T_1 \circ ... \circ T_t$. To that end, ATM procedure for map from regression described in \ref{sec:atm} is used to solve the problem:
\begin{equation*}
    {\mathcal{T}_t} = \underset{T}{\text{argmin }} \frac{1}{2} \sum_{i=1}^N (T(\mathbf{x}^i) - \mathbf{y}^i)^2,
\end{equation*}
where $\mathbf{x}^i \sim \rho$ and $\mathbf{y}^i = \left[T_1 \circ ... \circ T_{t} \right](\mathbf{x}^i)$.

This step of compression is usually performed each time the length of the composition of maps defining $\mathcal{T}_t$ reaches some maximum length assumed to be too expensive to evaluate. 

\subsubsection{Recovery maps}

One other drawback of using the composition of intermediate maps is that each intermediate maps is computed with a certain degree of approximation. Hence, map $\mathcal{T}_t$ suffers from accumulation of error committed when computing intermediate maps. 

One way to limit this accumulation of error is, at a given recovery time step $t$, to replace the composition of intermediate maps $\mathcal{T}_t$ by the direct posterior map $T_t$ such that:
\begin{equation}\label{eq:target_recovery}
    {T_t}_\sharp \rho = \pi(\bm{\theta}|\mathbf{y}_{1:t}) = \prod_{k=1}^t \pi(\mathbf{y}_t|\bm{\theta}).\pi(\bm{\theta})
\end{equation}
This recovery step can be automated by monitoring and setting a threshold on the accumulation of error via the variance diagnostic \eqref{eq:var_diag_discr} and trace diagnostic \eqref{eq:trace_diag_cont} computed between map induced density ${\mathcal{T}_t}_\sharp \rho(\bm{\theta})$ and posterior $\pi(\bm{\theta}|\mathbf{y}_{1:t})$.

It is worth noting that computing a recovery map leads also to map compression since map $\mathcal{T}_t$ at the end of the recovery step is one single direct map.

Algorithm \ref{alg:online_phase} summarizes steps of the online characterization of posterior densities which is model evaluations free and only requires evaluations of transport maps.
\begin{algorithm}
\SetKwFunction{length}{length}
\caption{Sequential characterization of posterior densities}\label{alg:online_phase}
\KwIn{Measurement $\mathbf{y}_t$ at time step $t$. map component $S_{\mathbf{y}_t}$ from Algorithm \ref{alg:offline_phase}.}

Compute $T_1$ such that
    ${T_1}_\sharp \rho(\bm{\theta}) = \pi(\bm{\theta}|\mathbf{y}_1)$

Set $\mathcal{T}_1=T_1$

Compute variance diagnostic \eqref{eq:var_diag_discr} $\hat{\epsilon}_\sigma({\mathcal{T}_1}_\sharp \rho,\pi(\mathbf{\theta}|\mathbf{y}_{1}))$

Compute trace diagnostic \eqref{eq:trace_diag_cont} $\hat{\epsilon}_{\text{trace}}(\pi(\mathbf{\theta}|\mathbf{y}_{1}),{\mathcal{T}_1}_\sharp \rho)$
    
\For{$t \in \{2,...,n_t\}$ }{
  \uIf{$\hat{\epsilon}_{\text{trace}}<\epsilon_{\text{trace}}^{\max}$ and $\hat{\epsilon}_\sigma<{\epsilon}_{\sigma}^{\max}$ }{
    Compute $T_t$ such that ${T_t}_\sharp \rho(\bm{\theta}) = \pi(\mathbf{y}_t|\bm{\theta})\rho(\bm{\theta})$
    
    Set $\mathcal{T}_t=\mathcal{T}_{t-1} \circ T_t$
    }
   \Else{ 
   Compute map $T$ such that ${T}_\sharp \rho = \pi(\bm{\theta}|\mathbf{y}_{1:t})$
   
   Set $\mathcal{T}_t = T$
    }

     \If{$\length\left[\mathcal{T}_t\right]>l^{\max}$}{
   Compute map $T$ via regression such that $T=\mathcal{T}_t$

   Reset $\mathcal{T}_t = T$
   }
   
    Compute variance diagnostic \eqref{eq:var_diag_discr} $\hat{\epsilon}_\sigma({\mathcal{T}_t}_\sharp \rho,\pi(\mathbf{\theta}|\mathbf{y}_{1:t}))$

    Compute trace diagnostic \eqref{eq:trace_diag_cont} $\hat{\epsilon}_{\text{trace}}(\pi(\mathbf{\theta}|\mathbf{y}_{1:t}),{\mathcal{T}_t}_\sharp \rho)$
 }
\end{algorithm}

In the proposed approach the measurement-free phase described Algorithm \ref{alg:offline_phase} is not constrained by real-time and can in theory be performed independently of the second phase. This is convenient when the forward model is expensive compared to the real-time constraints. However, fixing the proposal distribution $\pi_0$ in \eqref{eq:param_samps} can lead to less accurate computations if the posterior density is "far" from $\pi_0$. To improve approximation capabilities of offline maps with a fixed number of samples, $\pi_0$ can be chosen closer to the posterior that needs to be characterize. One simple way to do that is to choose $\pi_0$ as the posterior at a time step $t_0$ such that $t_0 < t$ when likelihood function $\pi(\mathbf{y}_t|\bm{\theta})$ needs to be approximated. This leads to asynchronous computation of the measurement-free and data-free phases. While this can easily be done in practice, without precise knowledge of computational resources, measurement time acquisition etc. it is difficult to propose a general method to how systematically update the proposal distribution in real time. In the following numerical experiments, unless stated otherwise, choice will be made to fix $\pi_0$ as the prior of the Bayesian problem.

\section{Numerical examples}\label{sec:results}

\begin{figure}[h]
    \centering
    \includegraphics[width=0.45\textwidth]{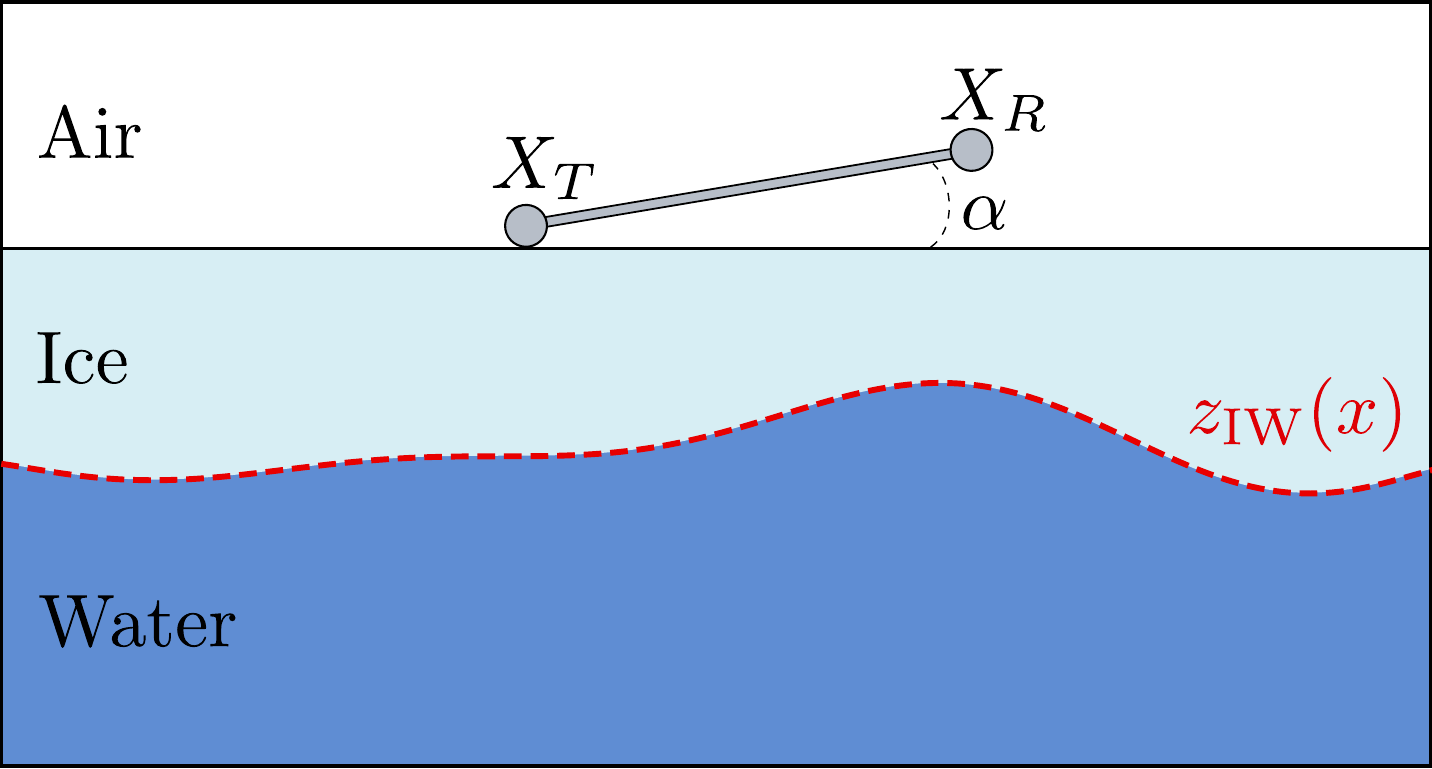}
    \caption{Problem geometry}
    \label{fig:geo}
\end{figure}

Numerical experiments in this section are in the context of sea ice thickness estimation from conductivity measurements coming from an EM31 device. Functioning of the EM31 device is described in \cite{em31}. The device is composed by an electromagnetic dipole composed by a transmitter $X_T$ and receiver $X_T$ (see Figure \ref{fig:geo}) which provides ground effective conductivity measurements. Assuming a bi-layer model for the ground composed by ice and water with known conductivities, the measurement of the effective conductivity can easily lead to the estimation of ice thickness. From the sequential measurement of effective conductivities along the top ice surface the objective is to characterize the ice-water interface $z_{IW}$. The angle $\alpha$ in Figure \ref{fig:geo} represents a position error of the EM31 device that could occur when taking measurement along the top ice surface. Parametrized models of increasing complexities for the interface will be considered and parameters will be updated with EM31 measurements via sequential Bayesian inference.

Modularity of this numerical example will be used to highlight properties of the algorithms designed in this work. 

\subsection{Analytical likelihood function}\label{sec:ex1}

In this first example the ice-water interface is assumed to be planar i.e. $z_{IW}(x)=\theta$, where $\theta$ will be the parameter of interest to describe the ice thickness. According to \cite{em31}, the relation between the measurement (effective conductivity $\sigma_\text{eff}$) and $h$ in that case is given by
\begin{equation}\label{eq:em31_analytic}
    \sigma_\text{eff}(\theta) = \sigma_I \left(1-R(\theta)\right) + \sigma_W R(\theta),
\end{equation}
with $R(\theta) = 1/\sqrt{4\theta^2+1}$ and where $\sigma_I$, $\sigma_W$ are ice and water conductivities.

By assuming an additive white Gaussian noise, the observation model between measurement and parameter of interest $\theta$ reads:
\begin{equation*}\label{eq:model_an}
    y =  \sigma_\text{eff}(\theta) + \epsilon, \,\,\, \epsilon \sim N(0,\sigma_\epsilon)
\end{equation*}

In that case and if there is no tilting angle ($\alpha = 0$) the closed form of the likelihood function for one measurement $y$ is known as:
\begin{equation}\label{eq:true_lkl}
    \pi(y|\theta) = \exp \left( -\frac{1}{2} \left(\frac{y - \sigma_\text{eff}(\theta)}{\sigma_\epsilon}\right)^2 \right)
\end{equation}

\begin{figure}[ht]
	\centering
	\subfigure[Error on a regular grid]{\label{fig:compare_log_lkl}
    \includegraphics{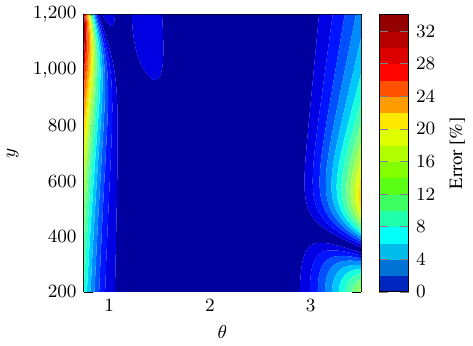}}
	\subfigure[Error at training samples]{\label{fig:compare_log_lkl_samp}
    \includegraphics{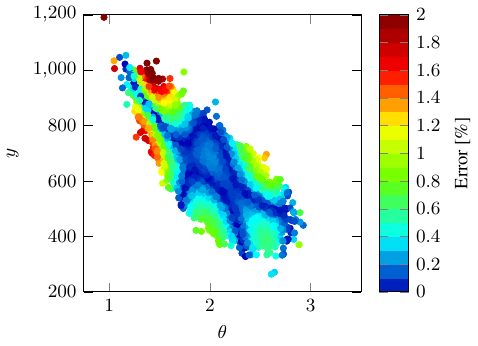}}
	\caption{Error between true and map-induced surrogate log-likelihood}
	\label{fig:compare_lkl}
\end{figure}

In the sequential setting, multiple measurements at different locations $t$ on the top ice surface are considered. Then, in each sequential posterior density $\pi(\theta|y_t)$, due to the problem translational symmetry, each likelihood function $\pi(y_t|\theta)$ (corresponding to each position $t$) follows the model \eqref{eq:true_lkl}. In this example likelihood ground truth is known and accuracy of the surrogate likelihood estimation of Algorithm \ref{alg:offline_phase} can be studied. Figure \ref{fig:compare_lkl} shows the error on the log-likelihood estimation via the transport map. The error is evaluated on a regular grid of both $\theta$ and $y$ i.e:
\begin{equation}
    \text{Error}(\theta,y) = \left| \frac{\log\pi(y|\theta)-\log\left[S(\theta;.)^\sharp \rho(y)\right]}{\log\pi(y|\theta)} \right|
\end{equation}
The two dimensional map $S$ used to represent the surrogate likelihood is computed with the ATM algorithm using 20,000 samples from the prior $\pi_0(\theta)=N(2,0.25)$ and corresponding outputs with a ten percent standard deviation noise ($\sigma_\epsilon=63$). Overall approximation of the likelihood via the computed transport map is very good, in most region of the parameter space the error is below two percents. Due to the under representation of training samples in that region, outside the range of four standard deviation ($\theta < 1$ and $\theta>3$) of the prior $\pi_0(\theta)$, approximation of the likelihood rapidly becomes less accurate. In the region where training samples are known, Figure \ref{fig:compare_log_lkl_samp} shows that the error is fairly small with some values in the top left corner around two percent and most of error being below one percent.

Based on the surrogate likelihood function, process described in Algorithm \ref{alg:online_phase} is applied to characterize sequential posterior densities. Measurements are simulated using the forward model with different noise realizations for each data assimilation step $t$ and a know reference value of parameters $\theta=2$. 

\begin{figure}[ht]
	\centering
 	\subfigure[Computation cost in terms of number of polynomial basis evaluations]{\label{fig:cost_reg}
     \includegraphics{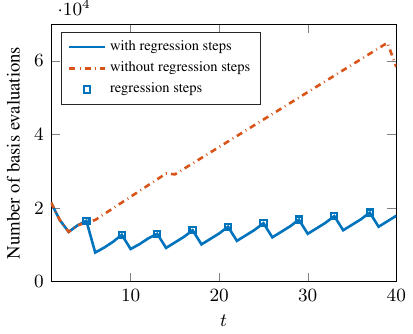}}
	\subfigure[Accuracy of posterior map approximations in terms of variance diagnostic]{\label{fig:var_diag_reg} 
    \includegraphics{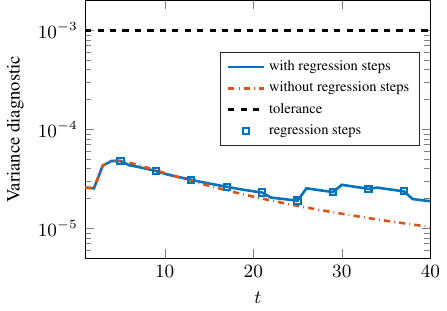}}
	\caption{Comparison of map qualities and cost with and without regression steps. Red dashed line represent computations where intermediate maps are sequentially composed to represent posterior densities. Blue solid line represent the case where, once the composition reach the length of five, a regression step is computed to find an unique map which approximate the composition of the five intermediate maps. Regression steps are represented by the square markers.}
	\label{fig:reg}
\end{figure}

Figure \ref{fig:reg} highlights the benefits of making regression steps. Regression steps are made each time the length of the composition of intermediate maps exceeds five. The maximum number of basis function (polynomial order) used to compute the map with the ATM algorithm is set to five. The stopping criterion (tolerance) in term of diagnostics are set to $\hat{\epsilon}_\sigma = 10^{-3}$ and $\hat{\epsilon}_\text{trace} = 10^{-2.5}$. Figure \ref{fig:cost_reg} compares the cost of the map characterizations in term of number of polynomial basis evaluations for each time (assimilation) step. This cost includes polynomial basis evaluations in the optimization of the posterior map but also the basis used to define the surrogate likelihood functions. Cost is high if one of more of this occur: the optimization requires a lot of iterations, the number of ATM iterations is high or the surrogate likelihood function is defined with a lot of basis functions. This is a good metric since in this model-free phase all computations are related to map evaluations  hence polynomial basis evaluations. This cost is also insensitive to hardware used to run the computations.  Without regression, the cost of posterior transport map computations linearly increase with a computation cost 
at time step 40 being three times larger than the cost for the first time step. Using recovery steps allows to keep a quasi-constant cost per step while keeping a good accuracy. Figure \ref{fig:var_diag_reg} shows the error of the transport maps characterization of posterior densities in terms of variance diagnostic. Simplifying composition of maps via regression doesn't affect the approximation too much and accuracy of map-based posterior densities approximation always stay below the tolerance threshold. Hence, by compromising a few approximation quality regression steps allow to have an online phase with a quasi-constant cost per iteration.

\begin{figure}[ht]
\centering
\subfigure[Computation cost in terms of number of polynomial basis evaluations]{\label{fig:cost_1d_reco}
\includegraphics{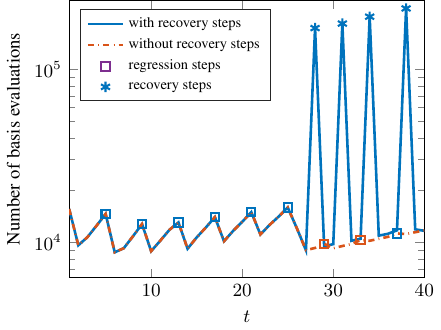}}
\subfigure[Accuracy of posterior map approximations in terms of variance diagnostic]{\label{fig:var_diag_1d_reco} 
\includegraphics{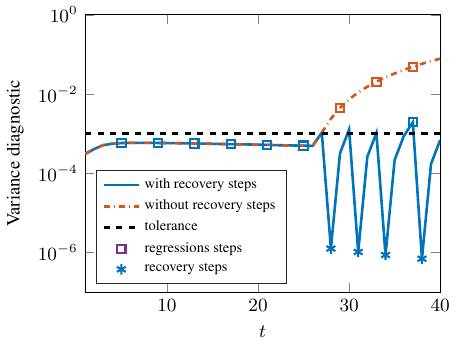}}
\caption{Comparison of posterior maps qualities and cost with and without recovery steps. Red dashed line represent computations where intermediate maps are sequentially composed to represent posterior densities. Blue solid line represent the case where, once the map accuracy exceed the tolerance, recovery steps are computed. In both cases regression steps (represented by square markers) are computed once the length of the composition exceeds five.}
\label{fig:reco}
\end{figure}

Figure \ref{fig:reco} illustrates the mechanic of the recovery steps. The problem to solve is the same than previously but the maps are now computed with less accuracy (by setting a larger tolerance on the computation of the intermediate maps). By doing that, as shown in Figure \ref{fig:cost_1d_reco}, map quality is very close to the tolerance in terms of variance diagnostic for the first 26 time steps. At time step 27, the variance diagnostic of the posterior map computation exceeds the tolerance which lead to computation of a recovery map at time step 28. After time step 28, still due to the coarser approximation of the intermediate maps, accumulation of error lead to parsimonious computations of recovery maps. For comparison, still in Figure \ref{fig:var_diag_1d_reco}, accuracy of the transport maps approximation without recovery steps is shown. Starting at time step 28, quality of posterior maps rapidly deteriorates without the recovery steps. Figure \ref{fig:cost_1d_reco} shows the counterparts of using the recovery maps: computation cost of recovery steps are about 20 times costlier than normal steps when intermediate maps are computed. This due to the fact that target densities in case of the computation of recovery maps \eqref{eq:target_recovery} are define via the product of likelihood functions and is more expensive than the target density for intermediate densities \eqref{eq:target_interm2}.

\begin{figure}[ht]
\centering
\includegraphics{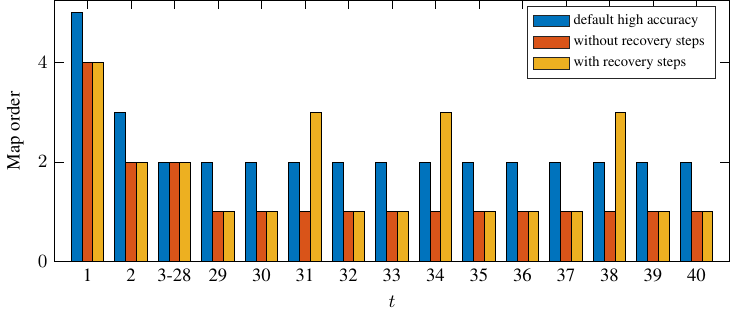}
\caption{Comparison of map complexity (map order) for each time step. Blue left bars represent the configuration presented in Figure \ref{fig:reg} where intermediate maps are computed with high accuracy. Red-middle and yellow-left bars represent the second configuration where intermediate maps are computed with lower accuracy. In the case represented by the yellow-left bars recovery maps are used.}
\label{fig:order_maps_1d}
\end{figure}

In Figure \ref{fig:order_maps_1d}, complexity of maps computed at each iterations is shown in terms of polynomial order. For each time steps the computed map is either an intermediate or recovery map. Left bars correspond to computations in the configuration of Figure \ref{fig:reg}, middle and right bars correspond to computations with and without recovery steps in the configuration of Figure \ref{fig:reco}. In the "high accuracy" setting intermediate maps are computed with higher complexity to satisfy a stricter tolerance. In the second configuration, where intermediate maps are computed less accurately, intermediate maps are less complex than in the previous case while recovery maps are in general more complex to compute (e.g., at time step 31, 34 and 38).

This comparisons show that, in general, it is more efficient to compute intermediate maps with the best accuracy possible. In the case where this is impossible due to limited online computational resources, diagnostics allow to know what is the level of the approximation errors and more expensive recovery maps can be computed at the next time step to recover these errors.

\begin{figure}[ht]
\centering
\includegraphics{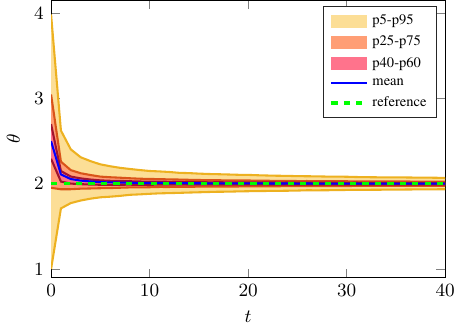}
\caption{Percentiles of the posterior densities at each time step}
\label{fig:prctiles_1d}
\end{figure}

Figure \ref{fig:prctiles_1d} shows percentiles of the posterior densities at each step. To compute those percentiles, at each step $t$ ($t=0$ correspond to the prior), the posterior map $\mathcal{T}_t$ are computed following Algorithm \ref{alg:online_phase} and posterior samples are generated as $\theta^i=\mathcal{T}_t(\epsilon^i)$ with $\epsilon^i \sim N(0,1)$. In this figure it can be observed that starting from a quite large off-centered prior, posteriors concentrates around the reference parameter value ($\theta=2$) used to simulate the measurements. 

\subsection{Case of nuisance parameters}
The following example is still in the context of ice-thickness characterization with the EM31 device. Hypotheses are complexified such that the ice-water interface is no longer assumed to be planar. In this context the analytical model described in \eqref{eq:em31_analytic} is no longer valid. To model the electromagnetic field inside the three-layers material the following eddy current model is used. This model known as a simplification of Maxwell's equations is suitable for describing the relatively low frequency field produced by the EM31. Let $H$ be a complex-valued magnetic field at a frequency of $w=0.98\text{kHz}$ and let $\sigma(x,z)$ be a spatially varying conductivity field, $H$ is assumed to be governed by the eddy current model:
\begin{equation}\label{eq:pde}
    \nabla^2 H - i \omega \sigma(x,z)H = 0.
\end{equation}

Effective conductivity $\sigma_\text{eff}$ measured by the EM31 device relates to the magnetic field $H$ by:
\begin{equation}\label{eq:conduc}
    \sigma_\text{eff} = \frac{4}{\omega \mu_0 \lVert X_T - X_R\rVert} \text{Im}\left[\frac{iH_I(X_R)}{H_R(X_R)} \right],
\end{equation}
where $X_T$ is the location of the transmission coil, $X_R$ is the location of the receiving coil (see Figure \ref{fig:geo}), $H=H_R+iH_I$, $\text{Im}[.]$ extracts the imaginary component, and $\mu_0\approx4\pi \times 10^{-7}$ is the permeability of free space. 

\begin{figure}[ht]
\centering
\includegraphics{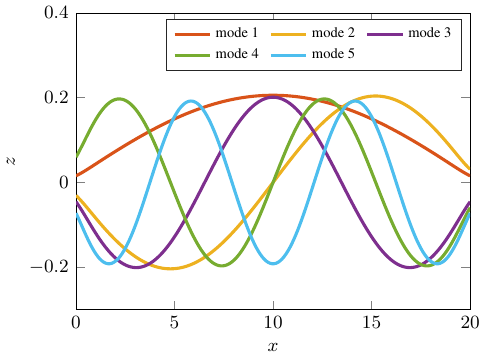}
\caption{Modes $\Psi_k$, $k=1,..,5$ used to parameterize the ice-water interface geometry}
\label{fig:kl_modes}
\end{figure}

\begin{figure}[ht]
\centering
\subfigure[H real component]{\label{fig:Hr} 
\includegraphics{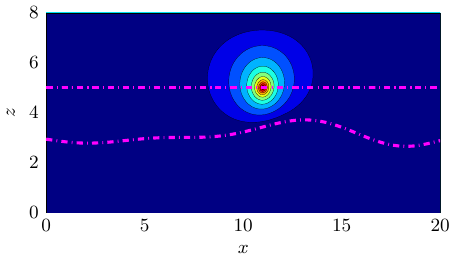}}
\subfigure[H imaginary component]{\label{fig:Hi}
\includegraphics{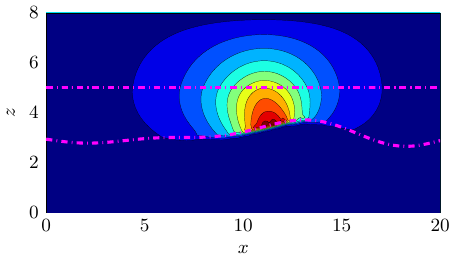}}
\caption{Eddy current solution. Dashed lines represent the interfaces between air, ice and water.}
\label{fig:eddy_current}
\end{figure}

In this case, parameters $\bm{\theta}$ used to characterize the ice-water interface enters in model with the conductivity field $\sigma$ which is defined by the geometry of the material interfaces. To parameterize the more complex ice-water interface a modal decomposition is used such that:
\begin{equation}\label{eq:model_interface}
    z_{IW}(x) = \theta_0 + \sum_{k=1}^m \theta_k \Psi_k(x),
\end{equation}
where $\Psi_k$ are the spatial modes defined in Figure \ref{fig:kl_modes}. 
For a particular choice of parameters the geometry of the ice-water interface can be defined and the resulting effective conductivity is found using the FENICS package \cite{fenics}. Solving of the pde by the finite element method takes about 1s for a given position of the measurement device on a 16 cores CPU at 3.60GHz. In this setting, relation between model output and parameters is no longer explicit and gradients with respect to parameters are unknown. For a specific non-planar ice-water interface, numerical solution of the eddy current equation is shown in Figure \ref{fig:eddy_current}.
In this example dependency of the magnetic field components with respect to the ice-water interface can be noticed which gives an insight of why the previous analytical model can't be used for non-planar interfaces.

In this example, it is also assumed that the EM31 positioning on the top ice surface is not perfect. More precisely it is assumed that the device can be tilted of angle $\alpha$ with respect to ground as illustrated in the Figure \ref{fig:geo}. This can occur, for example, when pulling the device from the right and/or if the terrain is rough. 

\begin{figure}[ht]
\centering
\subfigure[$t=0$ (prior)]{\label{fig:joint_KL1_0} 
\includegraphics{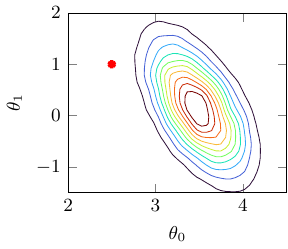}}
\subfigure[$t=10$]{\label{fig:joint_KL1_10}
\includegraphics{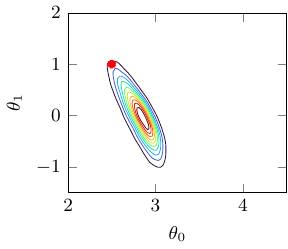}}
\subfigure[$t=20$]{\label{fig:joint_KL1_20}
\includegraphics{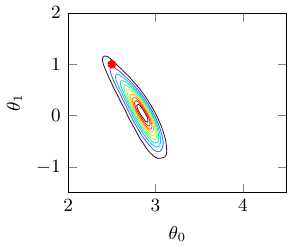}}
\subfigure[$t=30$]{\label{fig:joint_KL1_30} 
\includegraphics{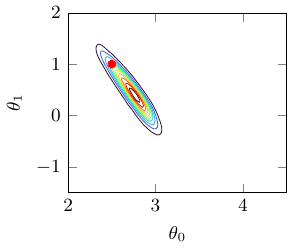}}
\subfigure[$t=40$]{\label{fig:joint_KL1_40}
\includegraphics{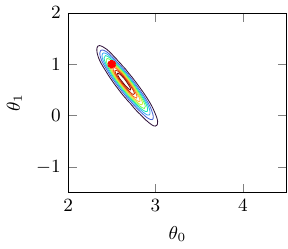}}
\subfigure[$t=50$]{\label{fig:joint_KL1_50}
\includegraphics{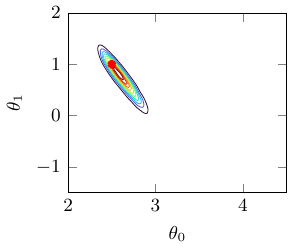}}
\caption{Joint posterior densities contours at different assimilation $t$. Red marker represent the reference $\bm{\theta}_\text{ref}$ used to simulate the observations}
\label{fig:joints_KL1}
\end{figure}

The tilt angle is assumed to be drawn independently from the uniform distribution on the $[0,5]$ interval for each measurement. In addition to this nuisance parameter, a small additive Gaussian white measurement noise ($\sigma_\epsilon=7$) is considered in order to emphasize the effect of the nuisance parameters. For this example the ice-water interface is assumed to be parameterized with one mode meaning that the parameters of interest are $\bm{\theta}=(\theta_0,\theta_1)$ of the interface model \eqref{eq:model_interface}. The prior density is chosen as a multivariate Gaussian which enforce the ice thickness to be positive and thinner than two meters with a confidence of 99\% (see Figure \ref{fig:joint_KL1_0}). 

To build the surrogate likelihood functions (which no longer have a explicit form), at each time step (location of the EM31 on the top ice surface), 20,000 samples of measurements and parameters are generated as described in Algorithm \ref{alg:offline_phase} with $\pi_0(\bm{\theta})$ chosen as the prior. For comparison purposes, likelihood functions without considering the tilt angle nuisance are also computed following the same procedure.

\begin{figure}[ht]
\centering
\subfigure{\label{fig:prctiles_notilt1} 
\includegraphics{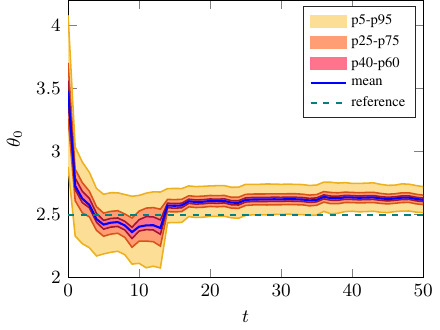}}
\subfigure{\label{fig:prctiles_notilt2}
\includegraphics{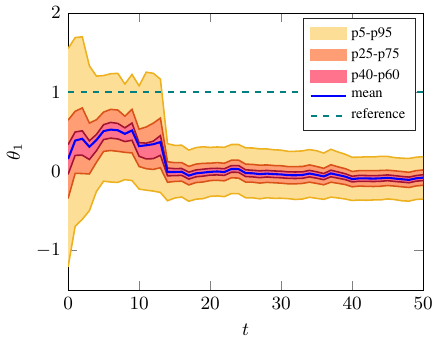}}
\caption{Percentiles and mean of marginals posterior densities when nuisance parameter is not considered (incorrect surrogate likelihood)}
\label{fig:prctiles_notilt}
\end{figure}

\begin{figure}[ht]
\centering
\subfigure{\label{fig:prctiles_tilt1} 
\includegraphics{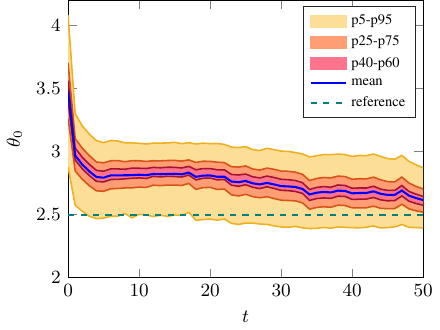}}
\subfigure{\label{fig:prctiles_tilt2}
\includegraphics{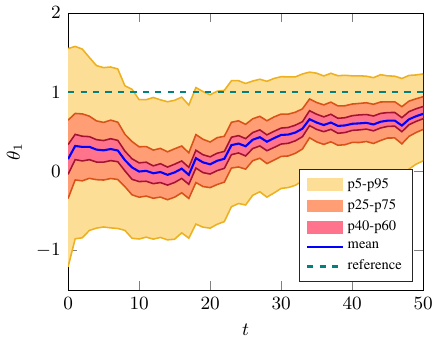}}
\caption{Percentiles and mean of marginals posterior densities when nuisance parameter is considered (correct surrogate likelihood)}
\label{fig:prctiles_tilt}
\end{figure}

Figures \ref{fig:prctiles_tilt} and  \ref{fig:prctiles_notilt} show percentiles of the marginal posteriors at each time step. In order to do that, posterior samples are generated by computing maps following Algorithm \ref{alg:online_phase} and the surrogate likelihood functions computed in the previous phase. To simulate observations, the numerical PDE-based model is evaluated at the reference parameter value $\bm{\theta}_\text{ref} = (2.5,1)$ and at realizations of nuisance parameters independently drawn from the uniform distribution on $[0,5]$ for each measurement location. In top of that measurement noise realizations are added to the model outputs.

Figure \ref{fig:prctiles_notilt} shows the result when the surrogate likelihood functions are "incorrect" meaning that it is computed without considering the effect of the nuisance parameters. Prediction of the marginal posterior marginal distribution do not converge nor include reference $\bm{\theta}_\text{ref}$ leading to wrong over-confident estimation of the parameters especially for parameter $\theta_1$. This show the effect of the nuisance parameter on the Bayesian solution and that a more complex likelihood model need to be used.

In Figure \ref{fig:prctiles_tilt}, solution using "correct" surrogate likelihood functions (computed considering nuisance parameters) is shown. As opposed to the previous case, despite the reference being in the tail of the prior, posterior mean converge to the reference and is almost always in the 90\% confidence interval. 
It is worth noting that with likelihood functions including dependence on the nuisance parameter leads to larger posteriors which take into account the additional uncertainty in the model.

\begin{figure}[ht]
\centering
\subfigure{\label{fig:var_KL1} 
\includegraphics{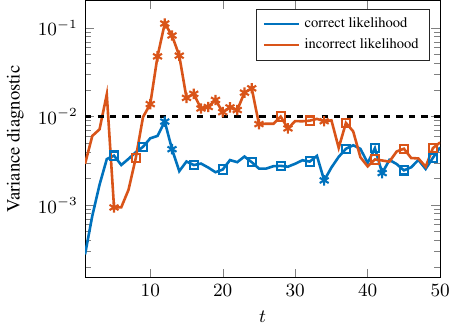}}
\subfigure{\label{fig:trace_KL1}
\includegraphics{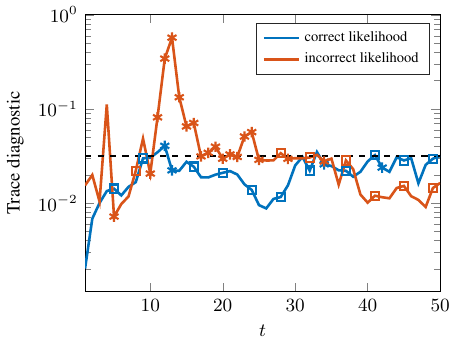}}
\caption{Error on the posterior maps in terms of variance and trace diagnostics for both cases of correct and incorrect surrogate likelihood functions. Asterisk markers correspond steps where recovery maps are computed while square markers correspond to steps where maps regressions are performed.}
\label{fig:diagnos_KL1}
\end{figure}

Figure \ref{fig:diagnos_KL1} shows the error committed on the estimation of posterior maps in terms of variance and trace diagnostics. It is clear that posterior maps are much harder to compute in the case of "incorrect" likelihood with a tolerance difficult to achieve leading to the computation of many recovery maps. This is due to the wrong construction of the Bayesian inference with incorrect likelihood functions which lead to singular densities. Further, since the posterior tries to "follow" the abnormal variation of the observations due to the presence of the nuisance, consecutive posteriors are not sufficiently close in order to allow the computation of intermediate maps to be efficient. When using the "correct" likelihood behavior of the proposed is similar to the example presented in Section \ref{sec:ex1} where most of the posterior maps are found via composition of intermediate maps with only few recovery maps computed (at steps 12,13,34 and 42). 
It is worth noting that for the "correct" likelihood case variance diagnostic tolerance is easier to satisfy than the trace diagnostic one. For the "incorrect" likelihood case, recovery steps are driven by both trace and variance diagnostics sometimes jointly, sometimes exclusively. 

\begin{figure}[ht]
\centering
\includegraphics{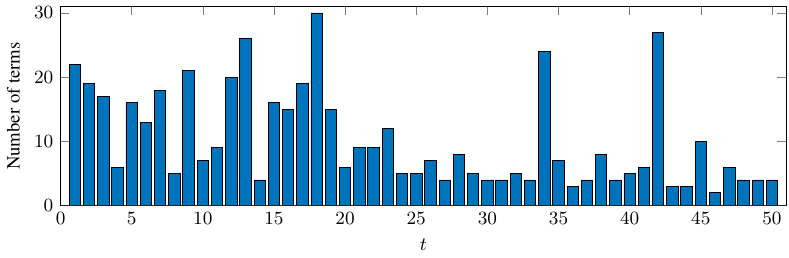}
\caption{Number of terms used basis used in the computation of the transport at each time step}
\label{fig:terms}
\end{figure}

\begin{figure}[ht]
\centering
\includegraphics{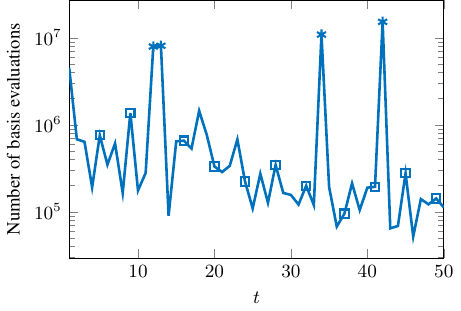}
\caption{Computation cost in terms of number of polynomial basis evaluations. Square markers correspond to regression steps and asterisk markers to recovery steps.}
\label{fig:cost_KL1}
\end{figure}

The joint two dimensional posterior densities for the estimation of ice-water interface in the presence of nuisance parameters are shown in Figure \ref{fig:joints_KL1} for some time steps. Non-Gaussian posterior densities can be observed for the first assimilation steps. 
These densities are represented by nonlinear posterior maps which can be verified in Figure \ref{fig:terms}. In this figure are shown the complexities of the maps computed at each step in terms of number of polynomial basis found by the ATM algorithm. For the first 20 steps maps are computed with higher complexity due to the higher complexity of the posterior densities and/or due to the more complex transformation between consecutive posterior densities.
Peaks of number of terms at steps 34 and 42 correspond to the computation of the recovery maps. Those peaks also clearly appear in Figure \ref{fig:cost_KL1} which show the computation cost of each step in terms of number of polynomial basis evaluation. It can be noticed that, as shown in the previous example, computation of recovery steps are always costlier than computation of intermediate maps. For example, despite the fact than map in step 18 is more complex to compute (30 terms), step 12 (computed with 20 terms) is more computational demanding due to the product of likelihood functions evaluated when computing the recovery map. Despite those parsimonious peaks of computation cost, the overall computation cost of the approach does not increase with the number of assimilation steps.

In this example where a nuisance parameter is considered, the non-Gaussian likelihood functions do not have an explicit form.
Hence, computing a reference solution in term of posterior densities is intractable. However, despite a reference value of parameters being set in a low probability region of the prior, the convergence of mean and other percentiles towards this value demonstrate the validity of the approach.

\subsection{Richer parametrization and comparison with MCMC}

In this example the complexity of the ice-water interface by increaseing using five modes of the geometry model \eqref{eq:model_interface}. This leads to the estimation of six parameters $\bm{\theta}=(\theta_0,...,\theta_5)$. Here the point is also to compare the approach to a standard MCMC approach. In order to do that, no nuisance parameter is considered leading to explicit likelihood functions:
\begin{equation}\label{eq:lkl_pde}
    \pi(y_t|\bm{\theta})=\exp \left( -\frac{1}{2} \left(\frac{y_t - \sigma_\text{eff}(\bm{\theta},t)}{\sigma_\epsilon}\right)^2 \right),
\end{equation}
where $\sigma_\text{eff}(\bm{\theta},t)$ represent the output of the model given by \eqref{eq:pde} and \eqref{eq:conduc} with conductivity field $\sigma(x,z)$ parameterize by $\bm{\theta}$ and transmitter at position $t$. 

\begin{figure}[ht]
    \centering
    \includegraphics{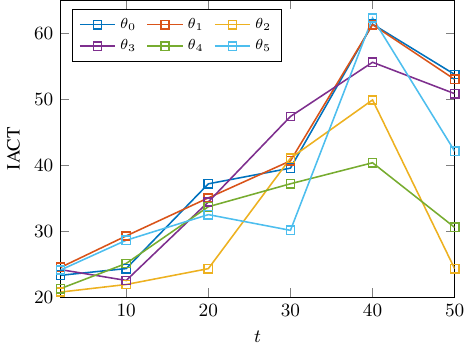}
    \caption{Integrated AutoCorrelation Times for each dimension and time step for the reference MCMC generated posterior samples at $t=2,10,20,30,40,50$. }
    \label{fig:IACTs}
\end{figure}

The transport-based solution is found thanks to the same procedure used before based on Algorithms \ref{alg:offline_phase} and \ref{alg:online_phase}. MCMC sampling is quite expensive with linearly increasing cost per time step. To draw one sample from the posterior at step $t$ with a MCMC algorithm, the numerical model needs to be evaluated $t-1$ times when $t>1$ due to the product of the likelihood functions in the sequential Bayesian formulation. Assuming that the posterior densities are unimodal we use the transport-based solution to define the starts of the MCMC chains (chosen as the mean of the transport-based solution). Assuming an already correct approximation by the transport map approach, following the proposal density of the adaptive Metropolis algorithm  \cite{Haario2001}, the proposal here is set to be a multivariate Gaussian with covariance matrix $(2.4^2/6).\Sigma_t$ where $\Sigma_t$ is the covariance of the posterior samples generated via the transport map approach. With those settings 100,000 MCMC samples are generated using a classical Metropolis-Hastings algorithm for $t=2,10,20,30,40,50$. In Figure \ref{fig:IACTs} are specified the Integrated AutoCorrelation Times (IACTs) for each dimension and step $t$.

\begin{figure}[ht]
\centering
\subfigure{\label{fig:pred_p0} 
\includegraphics{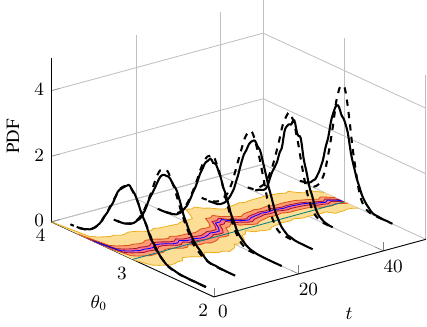}}
\subfigure{\label{fig:pred_p1}
\includegraphics{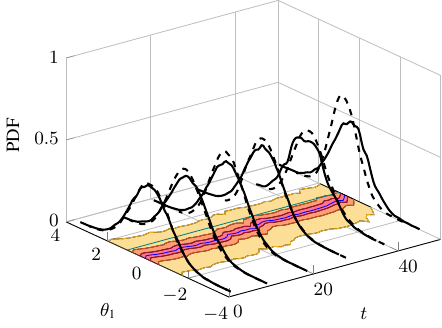}}
\subfigure{\label{fig:pred_p2}
\includegraphics{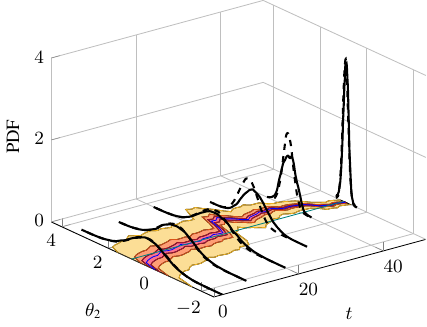}}
\subfigure{\label{fig:pred_p3}
\includegraphics{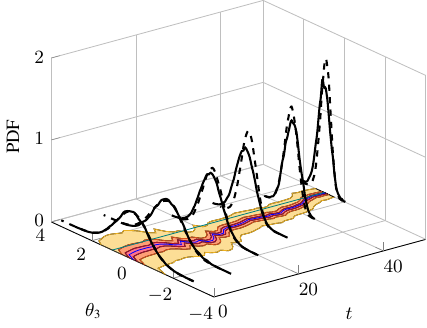}}
\subfigure{\label{fig:pred_p4}
\includegraphics{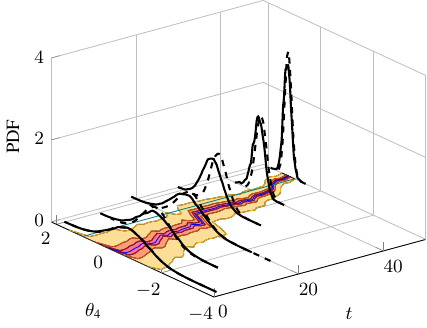}}
\subfigure{\label{fig:pred_p5}
\includegraphics{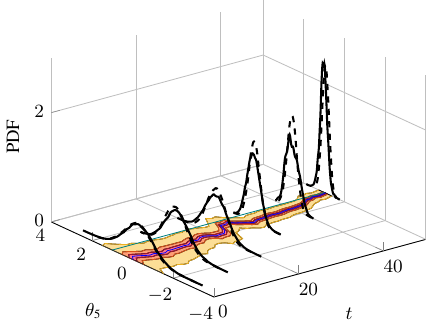}}
\subfigure{\label{fig:mcmcvstm_legend}
\includegraphics{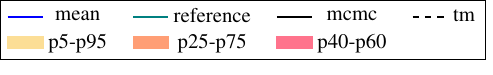}}
\caption{On the horizontal plane percentiles and mean of univariate marginals at each time step $t$. On the vertical axis MCMC (solid line) and transport map (dashed line) marginal densities are compared for $t=2,10,20,30,40,50$.}
\label{fig:quantile_mcmc_vs_tm}
\end{figure}

\begin{figure}[ht]
\centering
\subfigure[tm]{\label{fig:tm_2} 
\includegraphics[width=0.45\textwidth]{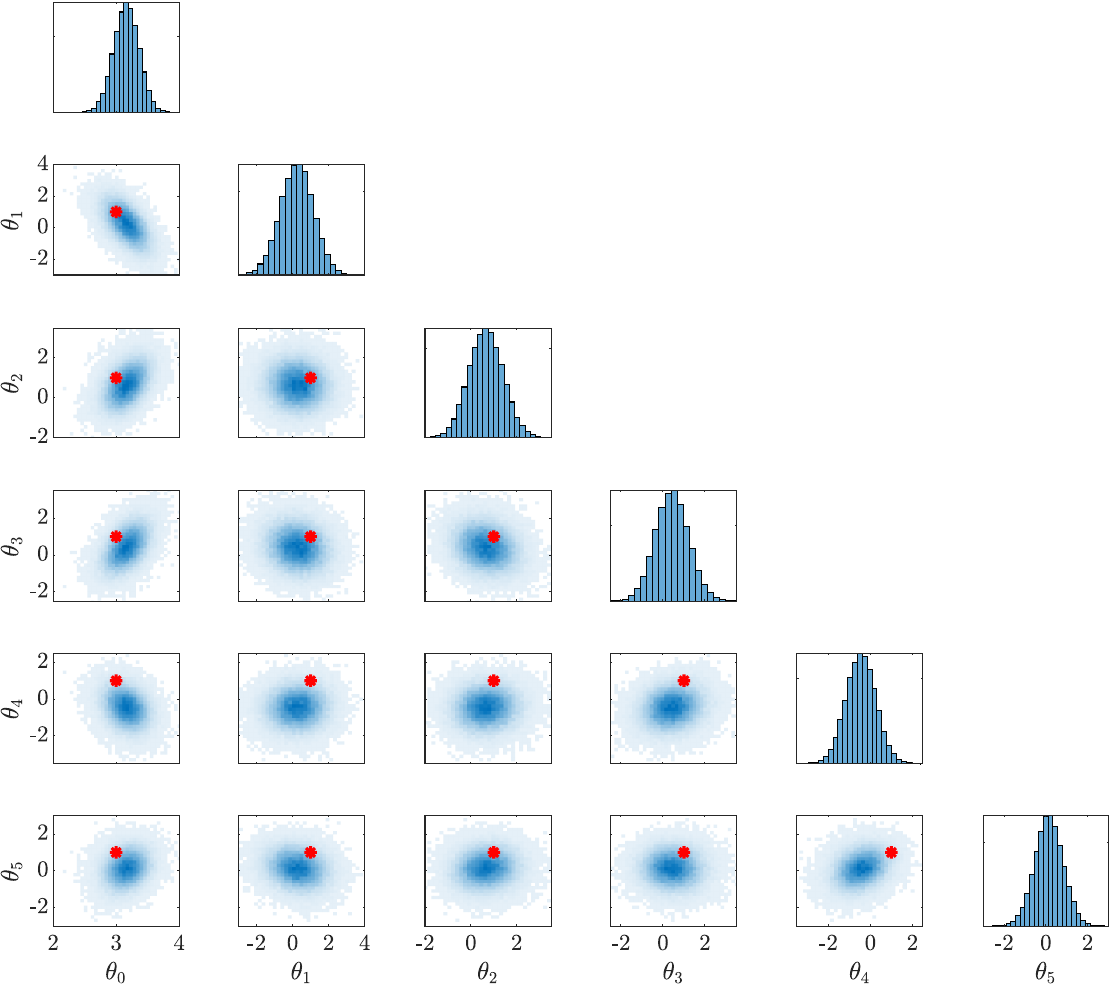}}
\subfigure[mcmc]{\label{fig:mcmc_2}
\includegraphics[width=0.45\textwidth]{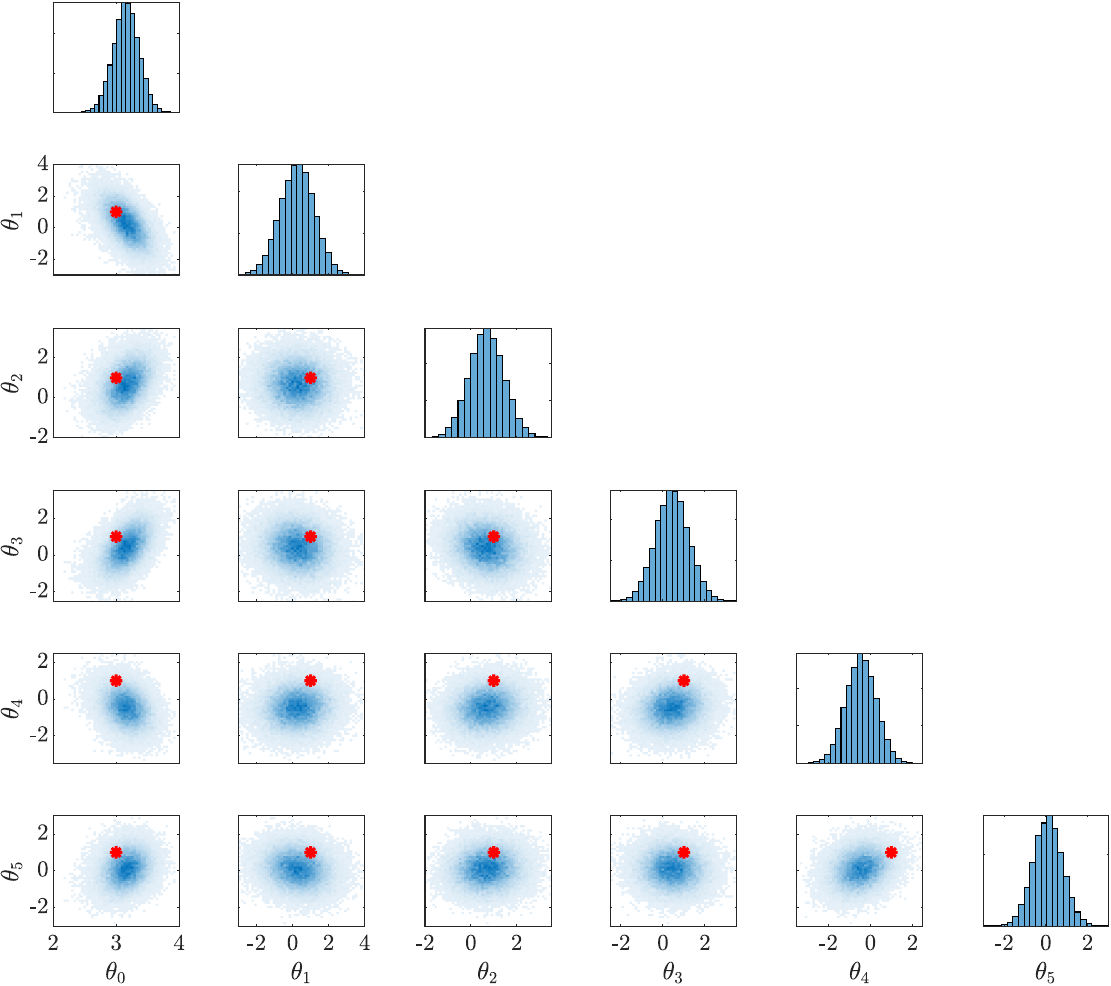}}
\caption{One and bi-dimensional histograms computed with MCMC and transport generated samples for $t=2$.}
\label{fig:samples_mcmc_vs_tm2}
\end{figure}

\begin{figure}[ht]
\centering
\subfigure[tm]{\label{fig:tm_50}
\includegraphics[width=0.45\textwidth]{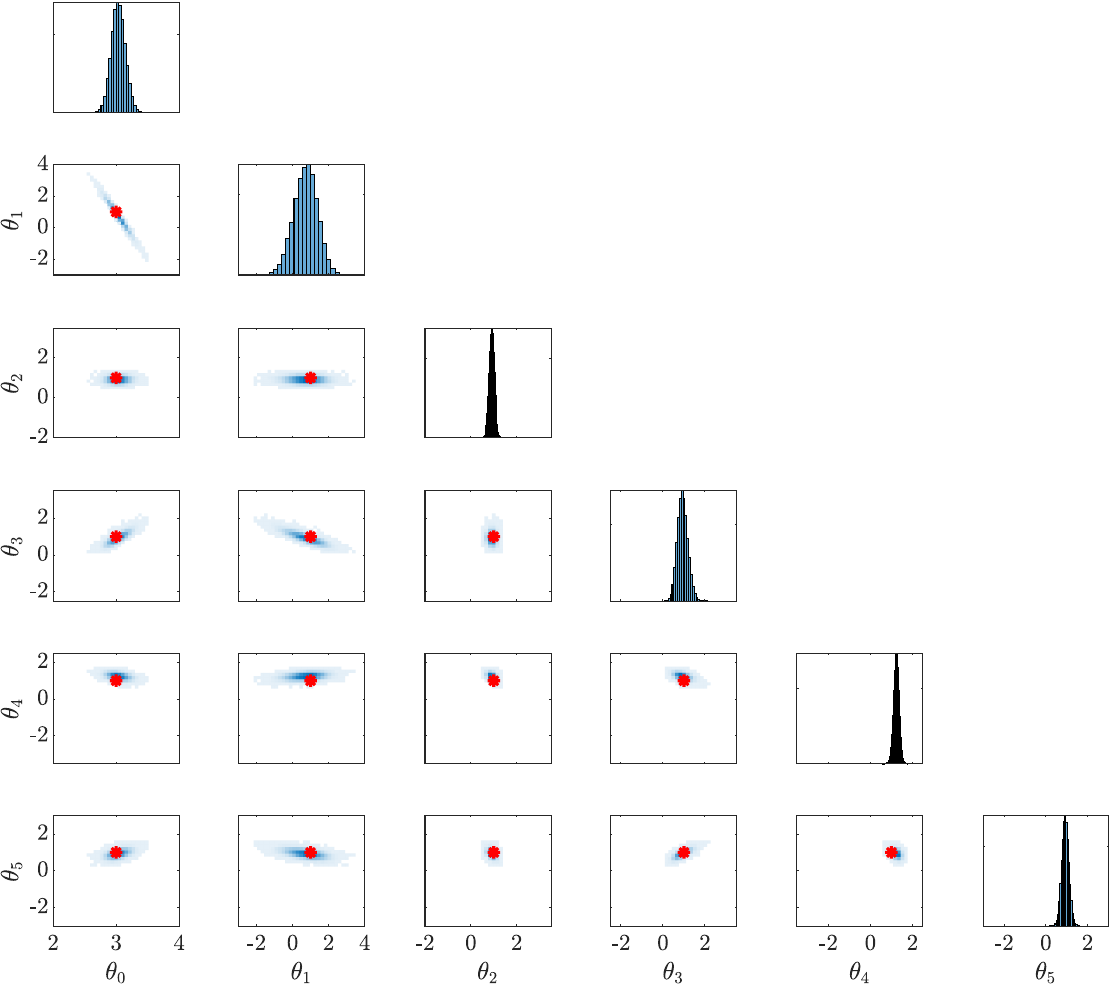}}
\subfigure[mcmc]{\label{fig:mcmc_50}
\includegraphics[width=0.45\textwidth]{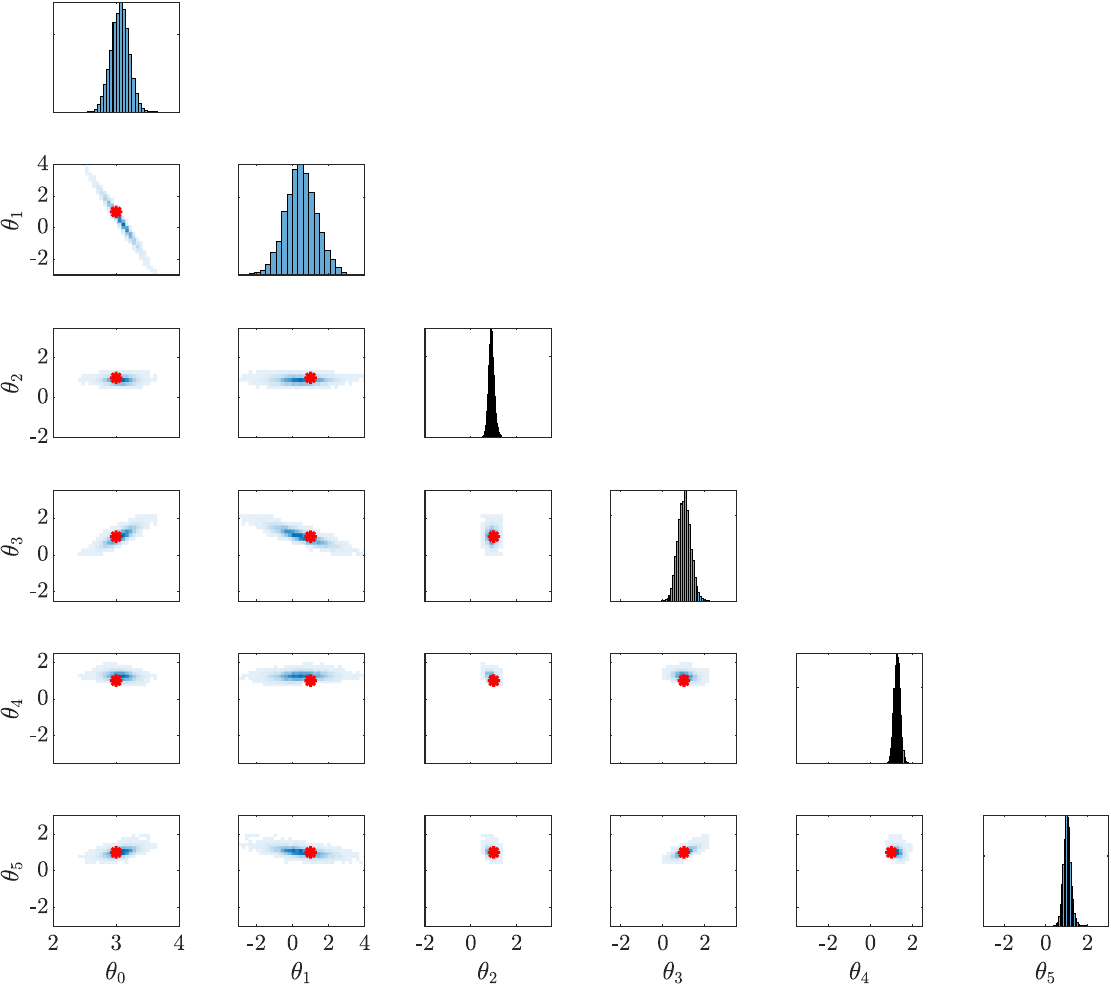}}
\caption{One and bi-dimensional histograms computed with MCMC and transport generated samples for $t=50$.}
\label{fig:samples_mcmc_vs_tm50}
\end{figure}

Figure \ref{fig:quantile_mcmc_vs_tm}, shows the solution of the sequential Bayesian inference problem in terms of percentiles and mean for the transport-based approach for each parameter. In top of that MCMC and transport-based posterior marginal densities computed via kernel density estimation are compared.
For each parameter, mean and percentiles computed with the transport-based algorithms converge towards the reference parameter values used to generate the synthetic observations. Despite some discrepancies both MCMC and the transport approach seem to agree. In the same spirit, comparison of 1D and 2D histograms are compared for posteriors at $t=2$ (Figure \ref{fig:samples_mcmc_vs_tm2}) and $t=50$ (Figure \ref{fig:samples_mcmc_vs_tm50}).
In those figures the reduction of the uncertainties with the assimilation of observations. For example, uncertainties on $\theta_2$ and $\theta_4$ are greatly reduced at $t=50$ compared to $\theta_1$. 

\begin{figure}[ht]
\centering
\subfigure[MCMC, $t=2$]{\label{fig:pred_mcmc2} 
\includegraphics{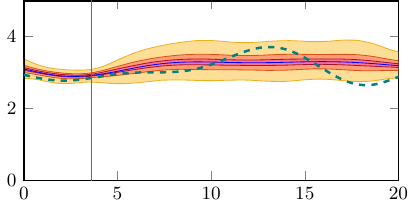}}
\subfigure[TM, $t=2$]{\label{fig:pred_tm2}
\includegraphics{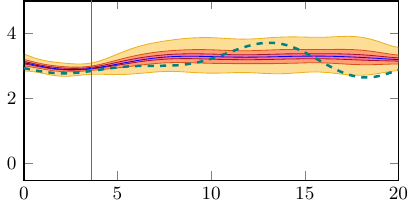}}
\subfigure[MCMC, $t=20$]{\label{fig:pred_mcmc20}
\includegraphics{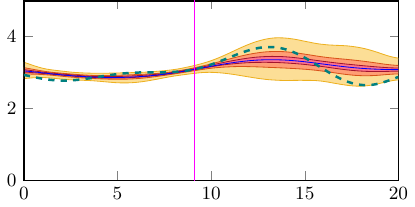}}
\subfigure[TM, $t=20$]{\label{fig:pred_tm20}
\includegraphics{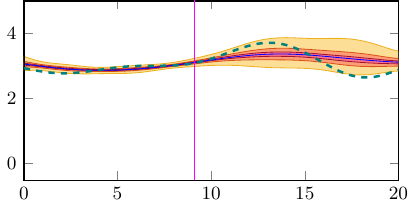}}
\subfigure[MCMC, $t=50$]{\label{fig:pred_mcmc50}
\includegraphics{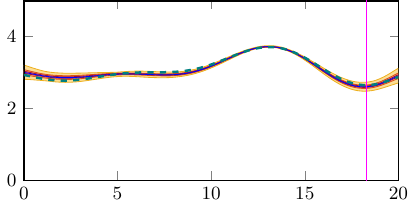}}
\subfigure[TM, $t=50$]{\label{fig:pred_tm50}
\includegraphics{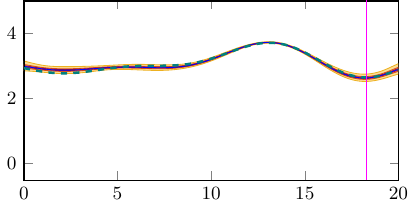}}
\subfigure{\label{fig:pred_legend}
\includegraphics{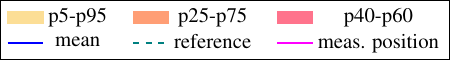}}
\caption{Percentiles and mean of the prediction of the ice-water interface at different time steps. Vertical cursor represent the position of the measurement device at each time step.}
\label{fig:pred_mcmc_vs_tm}
\end{figure}

With posterior samples, prediction of the ice-water can be done at each time step. In order to do that interface model \eqref{eq:model_interface} is evaluated at each posterior samples $\bm{\theta}^i$ and quantiles of corresponding output samples $z_{IW}^i(x)$ at some location $x$ can be computed. Figures \ref{fig:pred_mcmc_vs_tm} show predictions of ice-water in terms of percentiles and mean at $t=2,20,50$ using both transport and MCMC generated samples.
Predictions with both methods are very similar and the prediction at $t=50$ is a very good prediction of the truth (evaluated with reference value of parameters) with small confidence intervals. The interesting behavior of the assimilation problem which can be observed here is that the reduction of uncertainties follow the position of the measurement device. Indeed, despite having a global model for the ice-water interface, prediction is more concentrated for the geometry already "seen" by the measurement device (on the left of the vertical cursor in Figure \ref{fig:pred_mcmc_vs_tm}). Past geometry prediction is then lightly corrected with assimilation of further measurements.


\section{Conclusions}
The paper has presented a framework to solve sequential Bayesian inference problems for black-box models and complex likelihood models. The approach is non-intrusive for black box models and offer a flexible framework to tackle complex likelihood function models. The approach is based on two phases: (i) the observation-free phase where surrogate likelihood functions are built from model simulations (ii) the model-free phase where posterior densities are characterized. Both phases are based on the computation of monotone triangular maps which allow the characterization of the full Bayesian solution and the sampling of the possible non-Gaussian posterior densities. Use of accuracy diagnostics and the adaptive transport map framework allows efficient computation of maps given the information and computational resources available. In addition, monitoring of the error and computation of recovery maps allow to recover error at any step of the sequential process. The cost of the online phase only depends on the cost of polynomial basis evaluation and doesn't increase with the number of assimilation steps. In the numerical applications, the proposed approach lead to efficient characterization of unknown parameters even in the context of marginalized likelihood functions with nuisance parameters. While errors of the model-free phase can be quantified, error committed in the data-free can't be. Accuracy of the surrogate likelihood functions built in this phase relies on the number of joint parameter/data samples in the region of high posterior probability. Since this phase can be performed offline, high computational resources can be allocated to the generation of the training samples. An other approach not illustrated in this work is to perform the two phases asynchronously in order to use samples drawn according to intermediate posteriors instead of fixing the sampling distribution as the prior. This can easily be done since, at each time step, posterior samples are standard normal samples pushed by the posterior transport map computed in the online phase. Optimization of the communication between the phases depends on the model cost, computation resources etc. is left to do for specific applications.

One future direction is a more systematic treatments of nuisance parameters for sequential Bayesian inference problem. In this work focus has been made on nuisance parameters verifying hypothesis \ref{hyp:nuisance}. In this case likelihood functions at each time step are independent and the nuisance parameters act like a non-additive noise. In the general case where the realization of the nuisance parameter affects the prediction of the model at other time steps full likelihood $\pi(\mathbf{y}_{1:t}|\bm{\theta})$ needs to be estimated. A similar method that the one proposed in Algorithm \ref{alg:offline_phase} could be use. Difficulty of this will be the dimension of the densities that needs to be characterize which grows with the number of time steps. To keep reasonable and non increasing computation time sliding window dependence type hypothesis could be assumed in order to do the approximation: $\pi(\mathbf{y}_{1:t}|\bm{\theta})\approx\pi(\mathbf{y}_{1:t-\tau}|\bm{\theta})$ where $\tau$ is the maximum size of the sliding window. 

One target application of this approach could be parameter estimation in state-space dynamical models. A standard dynamical system reads:
\begin{equation*}
\left\{
\begin{array}{ll}
\mathbf{x}_{t} &=  F(\mathbf{x}_{t-1},\boldsymbol{\theta}) + \boldsymbol{\eta} \\
        \mathbf{y}_{t} &=  H(\mathbf{x}_{t}) + \boldsymbol{\epsilon} 
\end{array}
\right.
\end{equation*}
where $\mathbf{x}_{t}$ is the state at time step $t$, $\mathbf{y}_{t}$ is the observation at time step $t$, $F$ and $H$ are forward and observation operators respectively, $\boldsymbol{\eta}$ and $\boldsymbol{\epsilon}$ are model error and observation noise. If the objective is to characterize posteriors $\pi(\bm{\theta}|\mathbf{y}_{1:t})$ model error acts as a coupling nuisance parameter. It is worth noting that, in the case of no model error and uncertainty on the initial state in the dynamical system, the approach presented in this paper can directly be applied by simulating trajectories of the system and keeping joint samples of parameters and data. 

The presented approach provides an online phase which only require evaluation of transport maps (hence polynomial basis). By improving the computation efficiency of such object will directly affect the online computation cost and hence will improve the capability of doing real-time computation. Recent work software development \cite{MParT} aims toward this direction. 

\clearpage
\bibliographystyle{abbrv}

\begin{thebibliography}{1}

\bibitem{fenics}
M.~Aln{\ae}s, J.~Blechta, J.~Hake, A.~Johansson, B.~Kehlet, A.~Logg,
    C.~Richardson, J.~Ring, M.~E. Rognes, and G.~N. Wells.
\newblock {The FEniCS Project Version 1.5}.
\newblock {\em Archive of Numerical Software}, 3(100), 12 2015.

\bibitem{AndrieuPMCMC}
C.~Andrieu, A.~Doucet, and R.~Holenstein.
\newblock {Particle Markov chain Monte Carlo methods}.
\newblock {\em Journal of the Royal Statistical Society. Series B: Statistical
    Methodology}, 72(3), 2010.

\bibitem{Baptista2022LFI}
R.~Baptista, L.~Cao, J.~Chen, O.~Ghattas, F.~Li, Y.~M. Marzouk, and J.~T. Oden.
\newblock {Bayesian model calibration for block copolymer self-assembly:
    Likelihood-free inference and expected information gain computation via
    measure transport}.
\newblock 6 2022.

\bibitem{BaptistaATM}
R.~Baptista, Y.~Marzouk, and O.~Zahm.
\newblock {On the representation and learning of monotone triangular transport
    maps}.
\newblock 9 2022.

\bibitem{basu}
D.~Basu.
\newblock {On the Elimination of Nuisance Parameters}.
\newblock {\em Journal of the American Statistical Association}, 72(358):355, 6
    1977.

\bibitem{Beaumont2002}
M.~A. Beaumont, W.~Zhang, and D.~J. Balding.
\newblock {Approximate Bayesian Computation in Population Genetics}.
\newblock {\em Genetics}, 162(4):2025--2035, 12 2002.

\bibitem{Brennan2019}
M.~C. Brennan, D.~Bigoni, O.~Zahm, A.~Spantini, and Y.~Marzouk.
\newblock {Greedy inference with structure-exploiting lazy maps}.
\newblock 5 2019.

\bibitem{ChopinSMC2}
N.~Chopin, P.~E. Jacob, and O.~Papaspiliopoulos.
\newblock {SMC2: An efficient algorithm for sequential analysis of state space
    models}.
\newblock {\em Journal of the Royal Statistical Society. Series B: Statistical
    Methodology}, 75(3):397--426, 6 2013.

\bibitem{Cranmer2020}
K.~Cranmer, J.~Brehmer, and G.~Louppe.
\newblock {The frontier of simulation-based inference}.
\newblock {\em Proceedings of the National Academy of Sciences of the United
    States of America}, 117(48):30055--30062, 12 2020.

\bibitem{Cranmer2015}
K.~Cranmer, J.~Pavez, and G.~Louppe.
\newblock {Approximating Likelihood Ratios with Calibrated Discriminative
    Classifiers}.
\newblock 6 2015.

\bibitem{Rubin1984}
{Donald B. Rubin}.
\newblock {Bayesianly justifiable and relevant frequency calculations for the
    applied statistician}.
\newblock {\em The Annals of Statistics}, 12(4), 1984.

\bibitem{Moselhy2011}
T.~A. El~Moselhy and Y.~M. Marzouk.
\newblock {Bayesian inference with optimal maps}.
\newblock {\em Journal of Computational Physics}, 231(23):7815--7850, 9 2012.

\bibitem{Fearnhead2012}
P.~Fearnhead and D.~Prangle.
\newblock {Constructing summary statistics for approximate Bayesian
    computation: Semi-automatic approximate Bayesian computation}.
\newblock {\em Journal of the Royal Statistical Society. Series B: Statistical
    Methodology}, 74(3), 2012.

\bibitem{Frazier2022}
D.~T. Frazier, D.~J. Nott, C.~Drovandi, and R.~Kohn.
\newblock {Bayesian Inference Using Synthetic Likelihood: Asymptotics and
    Adjustments}.
\newblock {\em https://doi.org/10.1080/01621459.2022.2086132}, 2022.

\bibitem{Giraldi2014}
L.~Giraldi, D.~Liu, H.~G. Matthies, and A.~Nouy.
\newblock {To be or not to be intrusive? The solution of parametric and
    stochastic equations --- Proper Generalized Decomposition}.
\newblock 5 2014.

\bibitem{Haario2001}
H.~Haario, E.~Saksman, and J.~Tamminen.
\newblock {adaptive Metropolis algorithm}.
\newblock {\em Bernouilli}, 7(2):223--242, 2014.

\bibitem{Hermans2020}
J.~Hermans, V.~Begy, and G.~Louppe.
\newblock {Likelihood-free MCMC with amortized approximate ratio estimators}.
\newblock In {\em 37th International Conference on Machine Learning, ICML
    2020}, volume PartF168147-6, 2020.

\bibitem{Herzog2008}
M.~Herzog, A.~Gilg, M.~Paffrath, P.~Rentrop, and U.~Wever.
\newblock {Intrusive versus non-intrusive methods for stochastic finite
    elements}.
\newblock {\em From Nano to Space: Applied Mathematics Inspired by Roland
    Bulirsch}, pages 161--174, 2008.

\bibitem{DeepLearning}
Y.~LeCun, G.~Hinton, and Y.~Bengio.
\newblock {Deep learning (2015), Y. LeCun, Y. Bengio and G. Hinton}.
\newblock {\em Nature}, 521, 2015.

\bibitem{Lueckmann2017}
J.~M. Lueckmann, P.~J. Gon{\c{c}}alves, G.~Bassetto, K.~{\"{O}}cal,
    M.~Nonnenmacher, and J.~H. Mackey.
\newblock {Flexible statistical inference for mechanistic models of neural
    dynamics}.
\newblock In {\em Advances in Neural Information Processing Systems}, volume
    2017-December, 2017.

\bibitem{Marjoram2003}
P.~Marjoram, J.~Molitor, V.~Plagnol, and S.~Tavar{\'{e}}.
\newblock {Markov chain Monte Carlo without likelihoods}.
\newblock {\em Proceedings of the National Academy of Sciences of the United
    States of America}, 100(26), 2003.

\bibitem{em31}
J.~D. McNeill.
\newblock {Electromagnetic Terrain Conductivity Measurement at Low Induction
    Numbers}.
\newblock {\em Geonics Ltd.}, TN-6, 1980.

\bibitem{Papamakarios2016}
G.~Papamakarios and I.~Murray.
\newblock {Fast e-free inference of simulation models with Bayesian conditional
    density estimation}.
\newblock In {\em Advances in Neural Information Processing Systems}, 2016.

\bibitem{Papamakarios2018}
G.~Papamakarios, D.~C. Sterratt, and I.~Murray.
\newblock {Sequential Neural Likelihood: Fast Likelihood-free Inference with
    Autoregressive Flows}.
\newblock 5 2018.

\bibitem{MParT}
M.~Parno, P.-B. Rubio, D.~Sharp, M.~Brennan, R.~Baptista, H.~Bonart, and
    Y.~Marzouk.
\newblock {MParT: Monotone Parameterization Toolkit}.
\newblock {\em Journal of Open Source Software}, 7(80):4843, 12 2022.

\bibitem{Price2018}
L.~F. Price, C.~C. Drovandi, A.~Lee, and D.~J. Nott.
\newblock {Bayesian Synthetic Likelihood}.
\newblock {\em Journal of Computational and Graphical Statistics}, 27(1), 2018.

\bibitem{RubioControl}
P.-B. Rubio, L.~Chamoin, and F.~Louf.
\newblock {Real-time data assimilation and control on mechanical systems under
    uncertainties}.
\newblock {\em Advanced Modeling and Simulation in Engineering Sciences}, 8(1),
    2021.

\bibitem{Rubio2019}
P.-B. Rubio, F.~Louf, and L.~Chamoin.
\newblock {Transport Map sampling with PGD model reduction for fast dynamical
    Bayesian data assimilation}.
\newblock {\em International Journal for Numerical Methods in Engineering},
    2019.

\bibitem{Scanff2022}
R.~Scanff, D.~N{\'{e}}ron, P.~Ladev{\`{e}}ze, P.~Barabinot, F.~Cugnon, and
    J.~P. Delsemme.
\newblock {Weakly-invasive LATIN-PGD for solving time-dependent non-linear
    parametrized problems in solid mechanics}.
\newblock {\em Computer Methods in Applied Mechanics and Engineering},
    396:114999, 6 2022.

\bibitem{Spantini2019CouplingFiltering}
A.~Spantini, R.~Baptista, and Y.~Marzouk.
\newblock {Coupling techniques for nonlinear ensemble filtering}.
\newblock 6 2019.

\bibitem{Spantini2017}
A.~Spantini, D.~Bigoni, and Y.~Marzouk.
\newblock {Inference via low-dimensional couplings}.
\newblock {\em Journal of Machine Learning Research}, 19, 3 2018.

\bibitem{Thomas2022}
O.~Thomas, R.~Dutta, J.~Corander, S.~Kaski, and M.~U. Gutmann.
\newblock {Likelihood-Free Inference by Ratio Estimation}.
\newblock {\em Bayesian Analysis}, 17(1), 2022.

\bibitem{Zhegal2022}
J.~Zeghal, F.~Lanusse, A.~Boucaud, B.~Remy, and E.~Aubourg.
\newblock {Neural Posterior Estimation with Differentiable Simulators}.
\newblock 7 2022.

\end{thebibliography}

\end{document}